\documentclass{jfm}
\usepackage{amsmath}
\usepackage{graphicx}
\usepackage{natbib}
\usepackage{psfrag}
\usepackage{color}

\ifCUPmtlplainloaded \else
  \checkfont{eurm10}
  \iffontfound
    \IfFileExists{upmath.sty}
      {\typeout{^^JFound AMS Euler Roman fonts on the system,
                   using the 'upmath' package.^^J}%
       \usepackage{upmath}}
      {\typeout{^^JFound AMS Euler Roman fonts on the system, but you
                   don't seem to have the}%
       \typeout{'upmath' package installed. jfm.cls can take advantage
                 of these fonts,^^Jif you use 'upmath' package.^^J}%
      }
  \else
  \fi
\fi

\ifCUPmtlplainloaded \else
  \checkfont{msam10}
  \iffontfound
    \IfFileExists{amssymb.sty}
      {\typeout{^^JFound AMS Symbol fonts on the system, using the
                'amssymb' package.^^J}%
       \usepackage{amssymb}%
         
         \let\geq=\geqslant
      }{}
  \fi
\fi

\ifCUPmtlplainloaded \else
  \IfFileExists{amsbsy.sty}
    {\typeout{^^JFound the 'amsbsy' package on the system, using it.^^J}%
     \usepackage{amsbsy}}
    {\providecommand\boldsymbol[1]{\mbox{\boldmath $##1$}}}
\fi

\newcommand{\dy}{\partial}

\newcommand{\ex}{\mathrm{e}}

\newcommand{\zi}{{\rm i}}

\newcommand{\grad}{\nabla}

\newcommand{\eb}{\boldsymbol{e}}

\newcommand{\ub}{\boldsymbol{u}}

\newcommand{\Bb}{{\boldsymbol{B}}}

\newcommand{\Fb}{{\boldsymbol{F}}}

\newcommand{\jb}{{\boldsymbol{j}}}

\newcommand{\cA}{c_{\mathsf{a}}}
\newcommand{\cR}{c_{\mathsf{ra}}}

\newcommand{\Ubar}{{\overline{U}}}
\newcommand{\Bbar}{{\overline{B}}}
\newcommand{\Jbar}{{\overline{J}}}
\newcommand{\Qbar}{{\overline{Q}}}

\newcommand{\qhat}{{\hat{q}}}
\newcommand{\jhat}{{\hat{j}}}

\definecolor{dark-green}{rgb}{0,0.5,0} 

\title[Instability mechanism for MHD shear instabilities]{Interacting vorticity
waves as an instability mechanism for MHD shear instabilities}

\author[E. Heifetz, J. Mak, J. Nycander and O. M. Umurhan] {E. Heifetz$^{1,2}$,
J. Mak$^1$\thanks{Email address for correspondence:
julian.c.l.mak@googlemail.com.}, J. Nycander$^2$ and O. M. Umurhan$^{3,4}$}

\affiliation{$^1$Department of Geophysics \& Planetary Sciences, Tel Aviv
University, Tel Aviv, 69978, Israel
\\[\affilskip]
$^2$Department of Meteorology, Stockholm University, Stockholm, Sweden
\\[\affilskip]
$^3$NASA Ames Research, Space Sciences Division, Mail Stop N-245-3, Moffett
Field, CA 94043
\\[\affilskip]
$^4$SETI Institute, 189 Bernardo Ave., Suite 100, Mountain View, CA 94043
}

\date{DATE UPDATED}
\pagerange{\pageref{firstpage}--\pageref{lastpage}}
\doi{S002211200100456X}

\begin{document}
\maketitle
\label{firstpage}
\begin{abstract}

The interacting vorticity wave formalism for shear flow instabilities is
extended here to the magnetohydrodynamic (MHD) setting, to provide a mechanistic
description for the stabilising and destabilising of shear instabilities by the
presence of a background magnetic field. The interpretation relies on local
vorticity anomalies inducing a non-local velocity field, resulting in
action-at-a-distance. It is shown here that the waves supported by the system
are able to propagate vorticity via the Lorentz force, and waves may interact;
existence of instability then rests upon whether the choice of basic state
allows for phase-locking and constructive interference of the vorticity waves
via mutual interaction. To substantiate this claim, we solve the instability
problem of two representative basic states, one where a background magnetic
field stabilises an unstable flow and the other where the field destabilises a
stable flow, and perform relevant analyses to show how this mechanism operates
in MHD.

\end{abstract}

\begin{keywords}
***do this on submission website***
\end{keywords}
 


\section{Introduction}

Shear flows are ubiquitous in fluid systems, and shear flow instability,
nonlinear development and its transition into turbulence remains an active area
of research to the present day. We focus here on is magnetohydrodynamic (MHD)
shear instabilities, relevant to astrophysical systems such as, for example, the
solar tachocline, the magnetopause, and atmospheres of hot exoplanets. In
particular, we are interested in the physical mechanisms leading to ideal
parallel shear instabilities in MHD.

The argument generally is that, in the presence of a background magnetic field
that has a component parallel to the background flow, fluid instabilities have
to do work to bend field lines, thus the presence of a background magnetic field
should be stabilising. This is generally found to be true in planar geometry
when the background magnetic field is uniform
\citep[e.g.,][]{Chandrasekhar-Stability, Biskamp-MHD}. However, this argument
does not account for the observed destabilisation of hydrodynamically stable
flows in the presence of spatially varying background magnetic fields
\citep[e.g.,][]{Stern63, Kent66, Kent68, ChenMorrison91, TatsunoDorland06,
Leconanet-et-al10}, or the fact that a uniform field can destabilise some
wavenumbers that are hydrodynamically stable \citep[e.g.,][]{Kent66, Kent68,
RayErshkovich83}. Further, in two-dimensional spherical geometry (no radial
motion), a flow close to the observed solar differential rotation is
hydrodynamically stable, but is destabilised by the presence of an azimuthal
background magnetic field varying in latitude \citep[e.g.,][]{GilmanFox97,
GilmanCally07}. To reconcile these contrasting effects, the negative-energy wave
interpretation \citep[e.g.,][]{Cairns79}, based on wave resonance and energetic
arguments, is sometimes invoked \citep[e.g.,][]{RudermanBelov10}. The aim here
is to provide an intuitive mechanistic interpretation that reconciles the
contrasting influences of the magnetic field on MHD shear instabilities.

\begin{figure}
\begin{center}
	\psfrag{+qr}[ll][ll][0.8]{$\color{red}+q$}
	\psfrag{-qr}[ll][ll][0.8]{$\color{red}-q$}
	\psfrag{+qb}[ll][ll][0.8]{$\color{blue}+q$}
	\psfrag{-qb}[ll][ll][0.8]{$\color{blue}-q$}
	\psfrag{uy}[ll][ll][0.8]{$\Ubar(y)$}
	\psfrag{dq<0}[ll][ll][0.8]{$\Delta\Qbar<0$}
	\psfrag{dq>0}[ll][ll][0.8]{$\Delta\Qbar>0$}
	\psfrag{0}[cc][cc][0.8]{$0$}
	\psfrag{pi/2}[cc][cc][0.8]{$\pi/2$}
	\psfrag{-pi/2}[cc][cc][0.8]{$-\pi/2$}
	\psfrag{pi}[cc][cc][0.8]{$\pi$}
	\psfrag{hindering}[cc][cc][0.8]{hindering}
	\psfrag{helping}[cc][cc][0.8]{helping}
	\psfrag{growing}[cc][cc][0.8]{growing}
	\psfrag{decaying}[cc][cc][0.8]{decaying}
	\includegraphics[width=\textwidth]{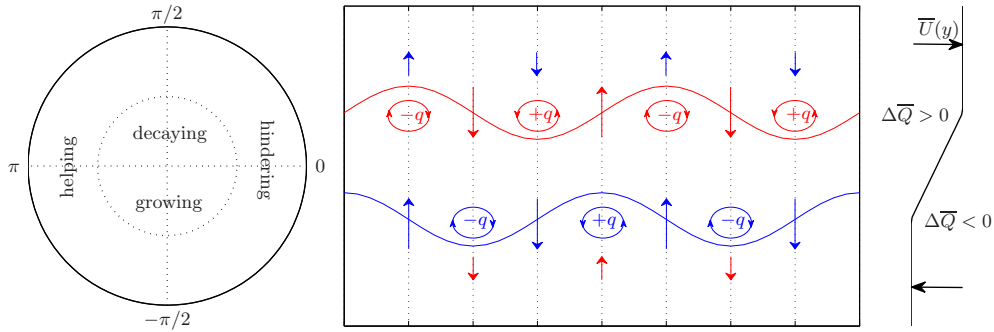}
	\caption{A pictorial representation of the wave interaction mechanism for the
	case of two Rossby waves. Here, $q\sim-\Delta\Qbar\eta$, where $q$ is
	vorticity, $\eta$ is displacement, and $\Delta\Qbar$ is the magnitude of the
	vorticity jump of the background flow profile. Taking into account the sign of
	$\Delta\Qbar$ at each jump, the $q$ anomalies resulting from the $\eta$
	distribution are labelled accordingly. The top wave has a self-induced
	propagation to the left relative to the mean flow, and vice versa for the
	bottom wave, and both waves propagating counter to the mean flow. The waves
	interact with the other via the non-local velocity field induced by local
	vorticity anomalies, and the induced velocities decay with distance from
	vorticity anomalies, represented here by the length of the arrows. The waves
	may be held steady owing to counter-propagation relative to the mean flow,
	together with the mutual interaction between the waves, resulting in a
	phase-locked configuration. The waves can then constructively interfere and
	lead to instability. A phase difference regime diagram is given on the left,
	where the phase-difference is defined as $\Delta\epsilon= \epsilon_{2} -
	\epsilon_{1}$, the lower wave displacement relative to the upper wave; the
	configuration here has $\Delta\epsilon=-\pi/2$.}
	\label{fig:CRW_basic}
\end{center}
\end{figure}

We present here an interpretation of shear instabilities in terms of interacting
vorticity waves. This interpretation includes as a special case the
Counter-propagating Rossby Waves (CRW) mechanism \citep[e.g.,][]{Bretherton66a,
Hoskins-et-al85}, a mechanism well-known in geophysical fluid dynamics, and has
been argued by \cite{BainesMitsudera94} to be an equivalent approach to the
negative-energy wave explanation for shear instability. A basic schematic of the
mechanism is recalled here for the Rossby wave case in
figure~\ref{fig:CRW_basic}, with a description of the mechanism given in the
caption.

The main ingredients required for instability in this interpretation are
phase-locking and constructive interference, achieved via counter-propagation of
waves relative to the mean flow, action-at-a-distance by the non-local velocity
field induced by local vorticity anomalies, and an appropriate phase-shift
between the two waves. Normal mode instability may also occur whenever the waves
are phase-locked with relative phase-difference (defined here in terms of wave
displacement) satisfying $\Delta\epsilon\in(-\pi,0)$. If the self-propagation
speed of each wave is large compare to its local mean flow, the
action-at-a-distance interaction should hinder the waves' propagation speed to
maintain phase-locking; this occurs when $\Delta\epsilon\in(-\pi/2,0)$, and
generally characterise long wavelength dynamics. By contrast, the
self-propagation speed of short wavelength waves counter to the mean flow is
generally weak, hence instability is obtained when the waves help each other's
propagation to overcome the background flow while growing, and this occurs when
$\Delta\epsilon\in(-\pi,-\pi/2)$. Essentially, the mechanism may be summarised
as \textit{``the induced velocity field of each Rossby wave keeps the other in
step, and makes the other grow''} \citep{Hoskins-et-al85}. For more details, we
refer the reader to the recent review of \cite{Carpenter-et-al13} and references
within.


In the case of figure~\ref{fig:CRW_basic}, we have illustrated the fundamental
mechanism using Rossby waves, but this is not the only possibility. As long as
we have wave modes that counter-propagate against the mean flow and they
propagate vorticity, it is perfectly possible for a configuration displayed in
figure~\ref{fig:CRW_basic} to occur. \cite{Harnik-et-al08},
\cite{Rabinovich-et-al11} and \cite{GuhaLawrence14} have shown that, in the
context of shear instabilities in stratified fluids, displacement of an
interface leads to an induced buoyancy field, that in turn induces an
appropriate vorticity field via baroclinic torque, thus Rossby--gravity waves
may also propagate vorticity. When two vorticity and/or density interfaces are
present, the interaction of the relevant wave modes leads to instability of the
Kelvin--Helmholtz type (two vorticity interfaces; e.g,
\citealt{DrazinReid-Stability}, \citealt{BainesMitsudera94}), Holmboe type
(essentially one vorticity and one density interface; \citealt{Holmboe62},
\citealt{BainesMitsudera94}), and the Taylor--Caulfield type (two density
interfaces; \citealt{Taylor31}, \citealt{Caulfield94}). For recent reviews and
studies of shear instabilities in stratified fluids and their interpretation in
terms of wave interactions, see \cite{Balmforth-et-al12},
\cite{Carpenter-et-al10, Carpenter-et-al13} and \cite{GuhaLawrence14}.

We show here that a similar interpretation holds in MHD. Wave displacement
changes the magnetic field configuration, and since the Lorentz force is
generally rotational, this in turn generates vorticity anomalies, so Alfv\'en
waves may propagate vorticity. We can thus use the vorticity wave interaction
framework as an interpretation for MHD shear instabilities. Furthermore, this
dynamical framework explains why we have stabilisation or destabilisation by the
magnetic field: the choice of basic state and parameter values affects the
properties of the wave modes, and the presence of instability depends on whether
supported wave modes can phase-lock and achieve mutual amplification.

Previous works on this topic have, for elucidation purposes, mainly considered
piecewise-linear basic states with interfacial wave solutions of the form
$q(x,y,t)=\qhat\delta(y-L)\ex^{\zi k(x-ct)}$, where $y=L$ is the location of
`jump' in the profile. If one can show that the vorticity generation is non-zero
only at these jumps, then interfacial wave solutions are exact solutions. A
small number of jumps in the basic state results in a dispersion relation that
is a low order algebraic equation with closed form solutions. Further analysis
of the solutions may be carried out in a relatively straight-forward manner
\citep[e.g.,][]{Rabinovich-et-al11}. One notable exception to this exactness in
the hydrodynamic setting is the Charney problem
\citep[e.g.,][]{Heifetz-et-al04b}, where the presence of differential rotation
$\beta$ results in non-localised vorticity generation. A similar phenomenon
occurs for the MHD case. Although this does not pose a problem for the numerical
computation of the eigenvalues and eigenfunctions, extra care is required when
interpreting the results and the physical mechanism. This subtlety is explained
in detail here, and the regimes where the interfacial wave assumption is a
reasonable approximation to the full solution are explored accordingly for the
instability problems we consider.

The layout of this article is follows. We provide the mathematical set up in
\S2, and explain how even simple Alfv\'en waves propagate vorticity via the
Lorentz force. Additionally, we demonstrate the non-local nature of vorticity
generation. For conceptual understanding of how waves in the system propagate
vorticity, we consider the dynamics of interfacial waves in \S3. To demonstrate
the instability mechanism, we solve the instability problem for two
piecewise-linear basic states, for comparison with previous work employing the
interacting vorticity wave formalism, and to test the performance of the
interfacial wave assumption. In \S4, we consider the case where a background
magnetic field stabilises the flow, taking the background flow to be the
Rayleigh profile demonstrated in figure~\ref{fig:CRW_basic}, together a uniform
background magnetic field. We first give details for the numerical method we
employ, then analyse in some detail the full solution, providing plots of
eigenfunctions and showing how the schematic in figure~\ref{fig:CRW_basic} is
modified by MHD effects. Analytic solutions resulting from the interface
assumption are derived, compared with the full solutions, and analysed
accordingly. We give a similar account in \S5 for the case where a linear shear
flow is destabilised by a spatially varying background magnetic field. We
conclude and discuss our results in \S6.


\section{Mathematical formulation}

The crux of the mechanism displayed in figure~\ref{fig:CRW_basic} is that waves
supported by the system and choice of basic state propagates vorticity, and
their interaction leads to instability. In this section, we provide the general
mathematical formulation, and explain how even simple Alfv\'en waves, supported
when the background flow and field are uniform, may propagate vorticity via the
Lorentz force. We consider the dynamics of waves when the flow and/or field is
sheared in the next section.


\subsection{Two dimensional MHD and action of Lorentz force}

We are interested here in ideal MHD instabilities. The homogeneous,
incompressible MHD equations are
\begin{subequations}\label{s2:full-equ}\begin{align}
	\frac{\dy\ub}{\dy t}+\ub\cdot\grad\ub&=-\frac{1}{\rho_{0}}\grad p
	+\frac{1}{\mu_{0}\rho_{0}}\jb^{*}\times\Bb^{*},\\
	\frac{\dy\Bb^{*}}{\dy t}+\ub\cdot\grad\Bb^{*}&=\Bb^{*}\cdot\grad\ub,\\
	\grad\cdot\ub=0,&\qquad \grad\cdot\Bb^{*}=0.
\end{align}\end{subequations}
Here, $\jb^{*}=\grad\times\Bb^{*}$ is the current. In Cartesian co-ordinates, an
analogue of Squire's theorem holds \citep[see e.g.,][and discussion
within]{HughesTobias01}, thus we may formulate the problem in two-dimensions.
The domain is taken to be the ($x,y$)-plane, with periodicity in $x$, and as yet
unspecified in $y$. The incompressibility condition allows us to write the
velocity $\ub$ and magnetic field $\Bb=\Bb^{*}/\sqrt{\mu_{0}\rho_{0}}$ (where
$\mu_{0}$ is the permeability of free-space and $\rho_{0}$ is the constant
density) in terms of a streamfunction $\psi$ and magnetic potential $A$, defined
here as $\ub=\eb_{z}\times\grad\psi$ and $\Bb=\eb_{z}\times\grad A$. From this,
the vorticity $q=\eb_{z}\cdot\grad\times\ub$ and current
$j=\eb_{z}\cdot\grad\times\Bb$ satisfy the relations $q=\grad^{2}\psi$ and
$j=\grad^{2}A$, and equations \eqref{s2:full-equ} take the equivalent form
\begin{equation}\label{s2:full-equ1}
	\frac{Dq}{Dt}=\grad\cdot(j\Bb),\qquad \frac{DA}{Dt}=0.
\end{equation}

\begin{figure}
\begin{center}
	\psfrag{j0}[ll][ll][0.8]{$j_{0}$}
	\psfrag{j1}[ll][ll][0.8]{$j_{1}$}
	\psfrag{F0}[ll][ll][0.8]{$F_{0}$}
	\psfrag{F1}[ll][ll][0.8]{$F_{1}$}
	\psfrag{-F0}[ll][ll][0.8]{$-F_{0}$}
	\psfrag{-F1}[ll][ll][0.8]{$-F_{1}$}
	\psfrag{Bx=B0>0}[ll][ll][0.8]{($a$) $B_{x}=B_{0}>0$}
	\psfrag{By=B0>0}[ll][ll][0.8]{($b$) $B_{y}=B_{0}>0$}
	\psfrag{+q}[cc][cc][0.8]{$+q$}
	\psfrag{djbx>0}[ll][ll][0.8]{$\cfrac{\dy}{\dy x}(jB_{x})>0$}
	\psfrag{djby>0}[ll][ll][0.8]{$\cfrac{\dy}{\dy y}(jB_{y})>0$}
	\includegraphics[width=\textwidth]{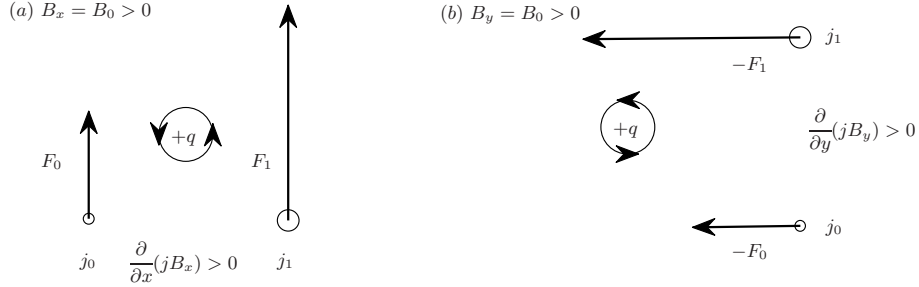}
	\caption{A pictorial representation of how the vorticity is generated by the
	Lorentz term $\jb\times\Bb=\grad\cdot(j\Bb)$ in two-dimensional
	incompressible MHD. For $j_{1}>j_{0}$, this generates the appropriately
	signed vorticity anomalies.}
	\label{fig:lorentz_force_vorticity}
\end{center}
\end{figure}

To see how the Lorentz force $\grad\cdot(j\Bb)$ results in rotational motion, we
suppose we have, for argument sake, $B_{0}\eb_{x}$ with
$B_{0}=\textnormal{const}>0$, and
\begin{equation}
	j(x_{0},y_{0})\eb_{z}=j_{0}\eb_{z},\qquad
	j(x_{1},y_{1})\eb_{z}=j_{1}\eb_{z},
\end{equation}
where $j_{1}>j_{0}$, $x_{1}>x_{0}$, $y_{1}=y_{0}$; this is a scenario where
$\dy(jb_{x})/\dy x>0$. Now, in general, the Lorentz term is $\Fb=\jb\times\Bb$,
and, with the right-hand-screw convention, the current distribution above
produces
\begin{equation}
	\Fb(x_{0},y_{0})=F_{0}\eb_{y},\qquad \Fb(x_{1},y_{1})=F_{1}\eb_{y},
\end{equation}
with $F_{1}>F_{0}>0$. A material line connecting $(x_{0},y_{0})$ and
$(x_{1},y_{1})$ is rotated anti-clockwise, i.e., this gives us a positive
vorticity anomaly; this is consistent with positive forcing in equation
\eqref{s2:full-equ1}. This is shown pictorially in
figure~\ref{fig:lorentz_force_vorticity}($a$).

Analogously, we could instead take $B_{0}\eb_{y}$, and the
above current distribution but taking $x_{1}=x_{0}$ and $y_{1}>y_{0}$ (together
with the analogous assumptions); this is the scenario where $\dy(jb_{y})/\dy
y>0$. This results in
\begin{equation}
	\Fb(x_{0},y_{0})=-F_{0}\eb_{y},\qquad \Fb(x_{1},y_{1})=-F_{1}\eb_{y},
\end{equation}
where $F_{1}>F_{0}$, and this may be seen to produce also anti-clockwise motion
of a material line. This is again consistent with equation \eqref{s2:full-equ1},
and is shown in figure~\ref{fig:lorentz_force_vorticity}($b$). Then, in the
linear case, the resulting action is a superposition of the above scenarios and
others like it.


\subsection{Linearisation and Alfv\'en wave dynamics}

We now linearise \eqref{s2:full-equ1} about a general basic state
$\Ubar(y)\eb_{x}$ and $\Bbar(y)\eb_{x}$ (and thus $\Qbar=-\dy\Ubar/\dy y$ and
$\Jbar=-\dy\Bbar/\dy y$), which results in the system of equations
\begin{subequations}\begin{align}
	\left(\frac{\dy}{\dy t}+\Ubar\frac{\dy}{\dy x}\right)q=&
	-\frac{\dy\Qbar}{\dy y}\frac{\dy\psi}{\dy x}+\Bbar\frac{\dy j}{\dy x}
	+\frac{\dy\Jbar}{\dy y}\frac{\dy A}{\dy x},\label{s2:q-dim}\\
	\left(\frac{\dy}{\dy t}+\Ubar\frac{\dy}{\dy x}\right)A=&
	\Bbar\frac{\dy\psi}{\dy x}.\label{s2:A-dim}
\end{align}\end{subequations}
The cross-stream displacement $\eta$ is given by
\begin{equation}\label{s2:eta-dim}
	v=\frac{\dy\psi}{\dy x}=
	\left(\frac{\dy}{\dy t}+\Ubar\frac{\dy}{\dy x}\right)\eta,
\end{equation}
and, substituting this into \eqref{s2:A-dim} and integrating results in the
relation
\begin{equation}\label{s2:a-eta-relation}
	A=\Bbar\eta+A_{0}(y),
\end{equation}
where we will take the non-advective contribution $A_{0}$ associated with
non-conservative effects to be zero for the rest of this work.

For simple Alfv\'en waves \citep[e.g.,][]{Biskamp-MHD}, we take the case
$\Ubar=0$, and $\Bbar=\textnormal{constant}>0$ without loss of generality. With
this, the governing equations become $\dy q/\dy t=\Bbar(\dy j/\dy x)$ and
$\dy\eta/\dy t=\dy\psi/\dy x$. We observe that $\eta\sim A$ from relation
\eqref{s2:a-eta-relation}, i.e., a positive displacement is correlated with a
positive $A$ anomaly. Furthermore, since $\grad^2 A=j$ and the Laplacian is a
negative definite operator, this implies the relation $\eta\sim-j$.

\begin{figure}
\begin{center}
	\psfrag{+q}[ll][ll][0.8]{$\color{blue}+q$}
	\psfrag{-q}[ll][ll][0.8]{$\color{blue}-q$}
	\psfrag{+eta-j}[ll][ll][0.8]{$(+\eta,-j)$}
	\psfrag{-eta+j}[ll][ll][0.8]{$(-\eta,+j)$}
	\psfrag{djbx>0}[ll][ll][0.8]{$\color{blue}\cfrac{\dy}{\dy x}(jb_{x})>0$}
	\psfrag{djbx<0}[ll][ll][0.8]{$\color{blue}\cfrac{\dy}{\dy x}(jb_{x})<0$}
	\psfrag{a-label}[ll][ll][0.8]{($a$)}
	\psfrag{b-label}[ll][ll][0.8]{($b$)}
	\psfrag{c-label}[ll][ll][0.8]{($c$)}
	\includegraphics[width=\textwidth]{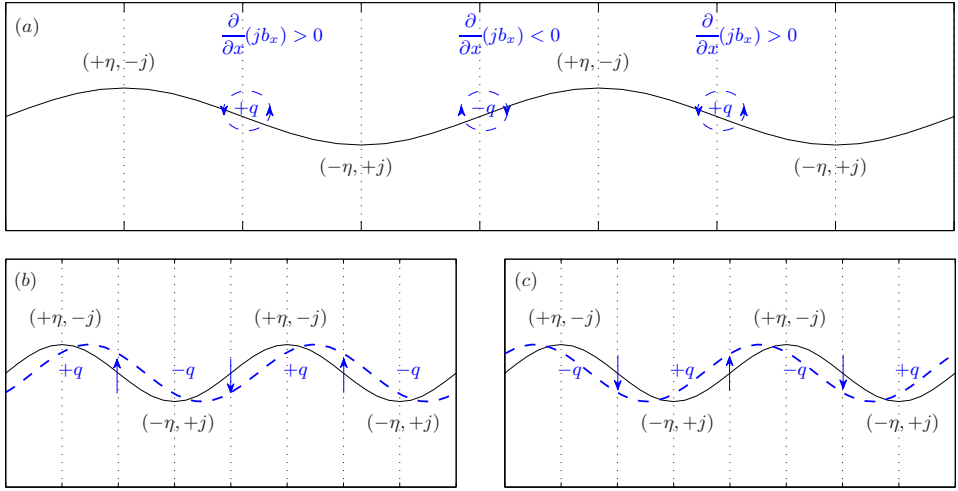}
	\caption{Pictorial representation for wave dynamics associated with
	$\Bbar(\dy j/\dy x)$, which is the linearised form of $\dy(jb_{x})/\dy x$,
	and includes simple Alfv\'en waves as a particular case. See text for
	description.}
	\label{fig:CRW_eigenstructure1}
\end{center}
\end{figure}

Suppose now we have some wave structure, as in
figure~\ref{fig:CRW_eigenstructure1}($a$). Associated with such a wave structure
is a $j$ distribution, anti-correlated with $\eta$ in this set up. From this,
since $\Bbar>0$, the distribution of $j$ leads to positive and negative
vorticity anomalies in between the wave crests and troughs depending on the sign
of $\dy j/\dy x$, resulting in the configuration in
figure~\ref{fig:CRW_eigenstructure1}($a$). To maintain a wave structure,
appropriate vorticity anomalies must exist in phase or anti-phase with $\eta$ to
counteract the action of the vorticity distribution at the notes. Alternatively,
the configuration in figure~\ref{fig:CRW_eigenstructure1}($a$) may be seen as a
superposition of equal amplitude right and left going waves \citep[e.g.,
appendix B of][]{Harnik-et-al08}. If $q$ is in phase with $\eta$ as in panel
($b$), then this gives a wave propagating to the right. On the other hand, if
$q$ is in anti-phase with $\eta$ as in panel ($c$), this gives the wave
propagating to the left. It may be checked from the resulting linearised
equations that, if we consider modal solutions of the form $\ex^{\zi k(x-ct)}$
(we take the $y$-wavenumber to be zero for simplicity), since the resulting
dispersion relation for simple Alfv\'en waves is $c^2=\pm\Bbar$ in this set up,
$c>0$ implies we have $\eta\sim-j\sim q$, whilst for $c<0$, $\eta\sim-j\sim-q$.
These relations are consistent with the schematic presented in
figure~\ref{fig:CRW_eigenstructure1}.

\subsection{Green's function formulation}

Wave interaction via action-at-a-distance may be represented mathematically by a
Green's function formalism, employed by previous authors
\citep[e.g.,][]{Harnik-et-al08}. Noting that $j=\grad^{2}A$ and
$\psi=\grad^{2}q$, and taking the domain to be $x$-periodic, a Fourier transform
leads to
\begin{equation}\label{s2:green}
	j=\left(-k^{2}+\frac{\dy^{2}}{\dy y^{2}}\right)A,\qquad
	q=\left(-k^{2}+\frac{\dy^{2}}{\dy y^{2}}\right)\psi,
\end{equation}
where $k$ denotes the $x$-wavenumber. \eqref{s2:green} may be formally inverted
to give
\begin{equation}\label{s2:inversion}
	\begin{pmatrix}A(y)\\ \psi(y)\end{pmatrix}=
	\int G(y,y')\begin{pmatrix}j(y')\\ q(y')\end{pmatrix}\ \mathrm{d}y',
\end{equation}
where $G(y,y')$ is a Green's function chosen to satisfy the boundary conditions.
Then, with the governing equations \eqref{s2:q-dim}, \eqref{s2:eta-dim}, and
substituting for the appropriate terms using relations \eqref{s2:a-eta-relation}
and \eqref{s2:inversion}, the problem is completely specified in terms of $q$
and $\eta$, where all intermediate effects from changes in $A$ and $j$ are
implicit.

For piecewise-linear profiles we consider in the subsequent sections, we
note this approach of using Green's functions guarantees the continuity of the
total pressure and displacement everywhere. The kinematic condition for the
continuity of the interface is already invoked in \eqref{s2:eta-dim}. From the
$y$-component of the linearised momentum equation, we observe that the inverted
$v$ and $b_{y}$ from $q$ and $j$ are guaranteed to be continuous by properties
of the Green's function, so the total pressure $p$ is also continuous since
$\Ubar$ and $\Bbar$ are continuous.


\subsection{Generation of vorticity away from interfaces}

If the background profiles are piecewise-linear, and, without loss of
generality, a `jump' is located at $y=L$, then $q\sim\qhat\delta(y-L)$ is exact
if it may be shown that vorticity generation is only non-zero at that location.
One notable exception however is the Charney problem,
\citep[e.g.,][]{Heifetz-et-al04b} where vorticity generation is non-local, and
we demonstrate here this occurs in the MHD case also. To show this, we take the
Laplacian of \eqref{s2:A-dim}, and the resulting linearised equation for $j$ is
given by
\begin{equation}\label{s2:j-dim}
	\left(\frac{\dy}{\dy t}+\Ubar\frac{\dy}{\dy x}\right)j=
	\left(\frac{\dy\Qbar}{\dy y}+2\Qbar\frac{\dy}{\dy y}\right)
	\frac{\dy A}{\dy x}
	+\Bbar\frac{\dy q}{\dy x}-\left(\frac{\dy\Jbar}{\dy y}+2\Jbar\frac{\dy}{\dy
	y}\right)\frac{\dy\psi}{\dy x}.
\end{equation}
Then, even if $(\dy\Qbar/\dy y, \dy\Jbar/\dy y) =(\Delta\Qbar,
\Delta\Jbar)\delta(y-L)$, $j$ and $q$ are not necessarily $\delta$-functions in
$y$ since $\Qbar(\dy A/\dy y)$ or $\Jbar(\dy\psi/\dy y)$ may not be zero for
$y\neq L$. Thus interfacial wave solutions fails to capture all the dynamics
since we are setting to zero contributions away from the interfaces, and we do
not know \emph{a priori} whether the neglected contributions are significant to
the dynamics. Further, non-local generation of vorticity implies that critical
layers will play a role in the dynamics, much like the Charney problem
\citep{Heifetz-et-al04b} and the second example considered in
\cite{Rabinovich-et-al11} for the stratified problem.

Although we will choose $q$ and $\eta$ as the fundamental variables for the bulk
of the discussion, this is not the only possibility. It turns out there are some
numerical advantages in utilising $q$ and $j$ as the variables when computing
for the full numerical solutions. The two choices are of course equivalent as
far as the full solution is concerned. A brief discussion of the ($q,j$)
equations is given in appendix~\ref{appendix2} to discuss how they differ when
interfacial wave dynamics are the focus.


\section{Interfacial wave dynamics}

We compute the wave modes supported in this MHD setting assuming a single
interface for conceptual progress, to see how we may expect waves to propagate
vorticity schematically. We consider an unbounded $y$-domain; the Green's
function for this setting is given by
\begin{equation}\label{s3:Greens-function}
	G(y,y')=-\frac{1}{2k}\ex^{-k|y-y'|}.
\end{equation}
We take piecewise-linear $\Ubar$ and $\Bbar$ (thus piecewise-constant $\Qbar$
and $\Jbar$), with
\begin{equation}
	\frac{\dy\Qbar}{\dy y}=\Delta\Qbar\delta(y-L),\qquad
	\frac{\dy\Jbar}{\dy y}=\Delta\Jbar\delta(y-L).
\end{equation}
Equations \eqref{s2:q-dim} and \eqref{s2:eta-dim} then become
\begin{subequations}\label{s3:equation1}\begin{align}
	\left(\frac{\dy}{\dy t}+\zi k\Ubar\right)q&=
	\zi k[-\Delta\Qbar\psi\delta(y-L)+\Bbar j+\Delta\Jbar A\delta(y-L)],\\
	\left(\frac{\dy}{\dy t}+\zi k\Ubar\right)\eta&=\zi k \psi.
\end{align}\end{subequations}
This set of resulting equations bears some formal resemblance to the analogous
stratified problem \citep[e.g.,][]{Harnik-et-al08}; this formal analogy is
detailed in appendix~\ref{appendix1}.

We consider solutions of the form
\begin{equation}
	q=\qhat\ex^{-\zi kct}\delta(y-L),\qquad
	j=\jhat\ex^{-\zi kct}\delta(y-L).
\end{equation}
With the inversion relation \eqref{s2:inversion}, the Green's function
\eqref{s3:Greens-function} for this domain, and taking modal solutions for
$\psi$ and $A$, we obtain the relations
\begin{equation}
	\psi(L)=-\frac{\hat{q}}{2k},\qquad
	A(L)=-\frac{\hat{j}}{2k}.
\end{equation}
Using $A=\Bbar\eta$ from \eqref{s2:a-eta-relation}, this results in
$\jhat=-2k\Bbar(L)\eta(L)$. Together, \eqref{s3:equation1}
becomes
\begin{equation}\label{s3:equation2}
	(\Ubar-c)\qhat=\frac{\Delta\Qbar}{2k}\qhat
	-2k\Bbar\left(\Bbar-\frac{\Delta\Jbar}{2k}\right)\eta,\qquad 
	(\Ubar-c)\eta=-\frac{1}{2k}\qhat,
\end{equation}
where all the functions associated with the basic state are taken to be
evaluated at $y=L$.

Now, combining the two equations in \eqref{s3:equation2} and solving the
resulting quadratic equation in $(\qhat/\eta)$, the eigenstructure and the
dispersion relation reads
\begin{equation}\label{s3:eigen-full}
	\qhat^{\pm}=2k(c^{\pm}-\Ubar)\eta^{\pm},\qquad
	(c^{\pm}-\Ubar)=-\frac{\Delta\Qbar}{4k}\pm
	\sqrt{\left(\frac{\Delta\Qbar}{4k}\right)^{2}
	+\Bbar\left(\Bbar-\frac{\Delta\Jbar}{2k}\right)}.
\end{equation}
We will call these modes generalised Rossby--Alfv\'en waves. The `plus' branch
is the one where the plus sign is taken, and analogously for the `minus' branch.
We observe that, when the waves are neutral, the plus branch is associated with
a wave propagating to the right, relative to the mean flow, and vice-versa for
the minus branch.


\subsection{A physical description of interfacial wave dynamics}

The eigenstructure \eqref{s3:eigen-full} tells us how $q$ is related to $\eta$
directly from the equations. To understand the relation in terms of changes in
the magnetic field configuration and how it results in vorticity anomalies and
wave propagation, it is informative to consider how the individual components
act. In \eqref{s3:eigen-full}, the presence of the $\Delta\Qbar$, $\Bbar^{2}$
and $\Bbar\Delta\Jbar$ terms is associated with the first, second and third term
on the right hand side of equation \eqref{s2:q-dim} respectively. The first term
of the three is the standard Rossby wave mechanism, for which the restoring
force comes from the background vorticity gradient. The wave propagation is that
already described in figure~\ref{fig:CRW_basic}. For the second term, this is
essentially the Alfv\'en wave case described in the previous subsection; the
eigenstructure relation \eqref{s3:eigen-full} may be seen to be consistent with
the schematic presented in figure~\ref{fig:CRW_eigenstructure1}.

\begin{figure}
\begin{center}
	\psfrag{+qb}[ll][ll][0.8]{$\color{blue}+q$}
	\psfrag{-qb}[ll][ll][0.8]{$\color{blue}-q$}
	\psfrag{+qr}[ll][ll][0.8]{$\color{red}+q$}
	\psfrag{-qr}[ll][ll][0.8]{$\color{red}-q$}
	\psfrag{+eta-j}[ll][ll][0.8]{$(+\eta,-j)$}
	\psfrag{-eta+j}[ll][ll][0.8]{$(-\eta,+j)$}
	\psfrag{by<0}[ll][ll][0.8]{$b_{y}<0$}
	\psfrag{by>0}[ll][ll][0.8]{$b_{y}>0$}
	\psfrag{dj<0}[ll][ll][0.8]{$\Delta\Jbar<0$}
	\psfrag{dj>0}[ll][ll][0.8]{$\Delta\Jbar>0$}
	\psfrag{djby>0}[ll][ll][0.8]{$\color{blue}\cfrac{\dy}{\dy y}(jb_{y})>0$}
	\psfrag{djby<0}[ll][ll][0.8]{$\color{blue}\cfrac{\dy}{\dy y}(jb_{y})<0$}
	\psfrag{by}[ll][ll][0.8]{$\Bbar(y)$}
	\psfrag{by=0}[cc][cc][0.8]{$\Bbar(y)=0$}
	\psfrag{a-label}[ll][ll][0.8]{($a$)}
	\psfrag{b-label}[ll][ll][0.8]{($b$)}
	\includegraphics[width=\textwidth]{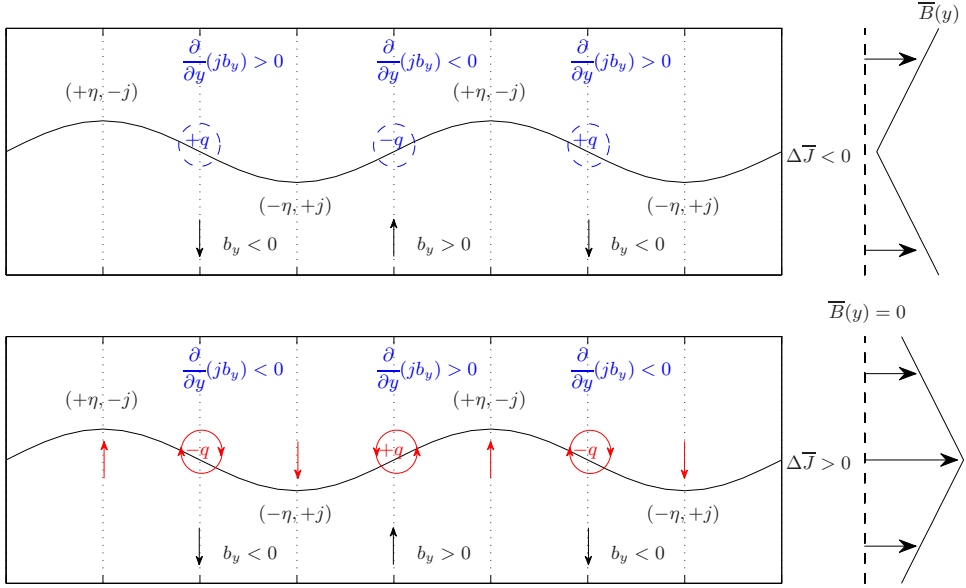}
	\caption{Pictorial representation for wave dynamics associated with
	$b_{y}(\dy\Jbar/\dy y)$, the third term on the right hand side of
	\eqref{s2:q-dim}, which is the linearised form of $\dy(jb_{y})/\dy y$.
	Taking $\Bbar>0$, we have again $\eta\sim-j$ as before (see paragraph just
	after equation \ref{s2:a-eta-relation}). Associated with the current
	anomalies is a distribution of $b_{y}$, given in black in both panels. Now,
	in ($a$), $\Delta\Jbar<0$, and so this gives the appropriate
	$\dy(jb_{y})/\dy y$ in blue, leading to vorticity anomalies depicted in
	blue. In this case, this is exactly analogous to the scenario depicted in
	figure~\ref{fig:CRW_eigenstructure1}, and leads to wave propagation. In
	($b$), $\Delta\Jbar>0$, and gives the appropriate vorticity distribution in
	red. In this case, there is no configuration of $q$ in phase or anti-pahse
	with $\eta$ that preserves the wave structure, and this leads to a
	prediction of an apparent local instability.}
	\label{fig:CRW_eigenstructure2}
\end{center}
\end{figure}

The third term on the right hand side of \eqref{s2:q-dim} is
\eqref{s2:q-dim} is
\begin{equation}
	\frac{\dy\Jbar}{\dy y}\frac{\dy A}{\dy x}=\frac{\dy\Jbar}{\dy y}b_{y}=
	\Bbar\frac{\dy\Jbar}{\dy y}\frac{\dy\eta}{\dy x},
\end{equation}
upon using the relation \eqref{s2:a-eta-relation}. We note $b_{y}(\dy\Jbar/\dy
y)$ is the linearised form of $\dy(jb_{y})/\dy y$. We show in
figure~\ref{fig:CRW_eigenstructure2} a schematic depiction of how the
$b_{y}(\dy\Jbar/\dy y)$ term acts, with the details given in the caption. From
the figure, we see that the sign of the vorticity anomalies generated depends on
the sign of $\Bbar\Delta\Jbar$. In the case where $\Bbar\Delta\Jbar<0$, we see
that both the second and third term in \eqref{s2:q-dim} generate a vorticity
distribution that contributes to wave propagation; see
figure~\ref{fig:CRW_eigenstructure1}($a$).

When $\Bbar\Delta\Jbar>0$, however, there is competition between the $\Bbar(\dy
j/\dy x)$ and $b_{y}(\dy\Jbar/\dy y)$ terms since they generate vorticity of the
opposite sign. The contribution from the latter term always dominates in the
long-wave limit, leading to the situation depicted in
figure~\ref{fig:CRW_eigenstructure2}($b$). This scenario suggests an instability
of the background magnetic field profile even in the absence of a background
flow. Although magneto-static profiles can suffer ideal MHD instabilities
\citep[e.g., Ch.~19,][]{GoldstonRutherford-Plasma}, we note here that, since we
are in planar geometry and we take $\Bbar(y)$, this basic state is stable in
ideal MHD by a theorem due to \cite{Lundquist51} because there the magnetic
field lines possesses no curvature. This prediction of instability at the local
level when the state is globally stable presumably stems from the fact that we
are considering isolated $\delta$-function solutions for the purposes of
elucidating the physics linking vorticity anomalies with wave displacement. The
neglected contributions and the resulting mutual interactions will presumably
suppress this apparent instability. For completeness, we note that, for the
hydrodynamic stratified setting, the schematic presented in
figure~\ref{fig:CRW_eigenstructure2}($b$) is formally similar to an
interpretation of Rayleigh--Taylor instability \citep[e.g.,][]{Harnik-et-al08}.

In the presence of a background shear flow, however, instability is possible. As
highlighted in \cite{Stern63}, taking a profile with $\Bbar\Delta\Jbar<0$
everywhere in the domain results in instability when the background flow is
stable in the hydrodynamic setting. Similar investigations in planar geometry of
the destabilising nature of the magnetic field have mainly considered profiles
where $\Bbar\Delta\Jbar<0$ in the domain, so here, for one of the examples, we
also take a basic state that satisfies this condition.


\section{Unstable profile stabilised by uniform magnetic field}

Having demonstrated how we expect waves to propagate vorticity anomalies, we now
consider two instability problems to demonstrate the instability mechanism, one
where the field stabilises an unstable flow, and the other where the field
destabilises a stable flow. Both basic states are chosen so that there is only
one non-dimensional parameter, given by $M=\tilde{B}/\tilde{U}$, where the
tildes denotes the relevant Alfv\'en speed and velocity scales. We note that,
from linear analysis, there is a stability theorem which states that
$|\Bbar|\geq|\Ubar|$ pointwise everywhere guarantees the absence of
exponentially growing instabilities \citep[e.g.,][]{HughesTobias01}; the basic
states are tailored so that the case $M\geq1$ is equivalent to the
aforementioned condition, with $M=1$ the case where we have equality.

We first provide details to the numerical method we employ to obtain our full
solutions, then analyse these solutions and explain how the instability
mechanism detailed in figure~\ref{fig:CRW_basic} is modified by MHD effects. To
test the validity of taking only a small number of interfaces, we consider the
approximated problem where only two interfaces are taken; closed form solutions
may be obtained from the resulting low order algebraic system, and these are
compared with the full solutions accordingly.


\subsection{Numerical method}

The governing equations depend on the choice of variables we employ for the
numerical scheme. If we describe the dynamics in terms of $q$ and $\eta$, the
governing equations are \eqref{s2:q-dim} and \eqref{s2:eta-dim}. Upon using
$j=\grad^{2}A=\grad^{2}(\Bbar\eta)$, where we have used the identity
\eqref{s2:a-eta-relation}, and with $\psi=\int q(y')G(y,y')\ \mathrm{d}y'$, the
resulting system of equations \eqref{s2:q-dim} and \eqref{s2:eta-dim} may be
written in discretised form as
\begin{equation}\label{s4:qeta-numerical}
	\frac{\dy}{\dy t}\begin{pmatrix}q \\ \eta\end{pmatrix}=
	-\zi k\begin{pmatrix} 
	\overline{\mathsf{U}}+\overline{\mathsf{Q}}'\mathsf{G} & 
	-\overline{\mathsf{B}}^{2}(-k^{2}\mathsf{I}+\mathsf{D}_{2})
	+2\overline{\mathsf{B}}\overline{\mathsf{J}}\mathsf{D}_{1} \\
	-\mathsf{G} & \overline{\mathsf{U}}\end{pmatrix}
	\begin{pmatrix}q \\ \eta\end{pmatrix},
\end{equation}
where, for example $\overline{\mathsf{U}}=\Ubar(y_{i})\times\mathsf{I}$, a prime
denotes a $y$-derivative, and $\mathsf{D}_{1,2}$ are the appropriate discretised
differential operators for $y$-derivatives. Taking uniform grid spacing $\Delta
y$, we take for example $\dy\Qbar/\dy y=\Delta\Qbar/\Delta y$ at the $y_{j}$
entry when $\dy\Qbar/\dy y=\Delta\Qbar\delta(y-y_{j})$. If a quantity $\Jbar$ is
discontinuous at $y_{j}$, we take
$\Jbar(y_{j})=[\Jbar(y_{j-1})+\Jbar(y_{j+1})]/2$. The discretised Green's
function is given by
\citep{Harnik-et-al08}
\begin{equation}
	\mathsf{G}_{m,n}=-\frac{\Delta y}{2k}\ex^{-k|y_{m}-y_{n}|},
\end{equation}
and this is a dense matrix. The system may be advanced in time accordingly if
one is considering an initial value problem in the context of non-modal
instabilities \citep[e.g.,][]{ConstantinouIoannou11}. For $(q,\eta)=
(\tilde{q},\tilde{\eta})\ex^{-\zi k(x-ct)}$, the eigenvalues and eigenvectors of
the system \eqref{s4:qeta-numerical} are the normal mode solutions.

It turns out it is numerically more stable to solve the problem in ($q,j$)
variables, thus we couple equation \eqref{s2:q-dim} with \eqref{s2:j-dim}
instead of \eqref{s2:eta-dim}. A similar manipulation using $A=\int
j(y')G(y,y')\ \mathrm{d}y'$ results in
\begin{equation}\label{s4:qj-numerical}
	\frac{\dy}{\dy t}\begin{pmatrix}q \\ j\end{pmatrix}=
	-\zi k\begin{pmatrix} 
	\overline{\mathsf{U}}+\overline{\mathsf{Q}}'\mathsf{G} & 
	-\overline{\mathsf{B}}-\overline{\mathsf{J}}'\mathsf{G} \\
	-\overline{\mathsf{B}}+\overline{\mathsf{J}}'\mathsf{G}
	+2\overline{\mathsf{J}}\mathsf{G}' 
	& \overline{\mathsf{U}}-\overline{\mathsf{Q}}'\mathsf{G}
	-2\overline{\mathsf{Q}}\mathsf{G}'
	\end{pmatrix}\begin{pmatrix}q \\ j\end{pmatrix},
\end{equation}
where, in this case, $\mathsf{G}'$ is the derivative of the Green's function,
with discretised form defined to be
\begin{equation}
	\mathsf{G}'_{m,n}=\begin{cases}
	+(\Delta y/2)\ex^{-k(y_{m}-y_{n})}, & y_{m}>y_{n},\\
	-(\Delta y/2)\ex^{+k(y_{m}-y_{n})}, & y_{m}<y_{n},\\
	0, & y_{m}=y_{n}.\end{cases}
\end{equation}
The value of $0$ is taken at $y_{m}=y_{n}$ because the wave supported on $y_{m}$
does not induce any $u$ or $b_{x}$ at $y_{m}$. The lack of derivative operators
in this latter matrix \eqref{s4:qj-numerical} results in a lower condition
number when compared to the ($q,\eta$) formulation in the test examples we have
considered. The two formulations are of course equivalent, and the relevant
eigenfunctions may be obtained from both formulations; here, the numerical
results presented were obtained from solving \eqref{s4:qj-numerical}.

Results presented here have been subject to domain size and resolution tests. A
domain size of $y\in[-5,5]$ was employed. A resolution of $\Delta y=10^{-2}$ (in
non-dimensional units) was found to be sufficient for modes away from
marginality, while for modes close to marginality, a resolution of $\Delta
y=5\times10^{-3}$ was employed instead to avoid the appearance of spurious
instabilities. The linear algebra problem was solved using the \verb|eig(A)|
command in MATLAB, and we pick out the mode with the largest imaginary part.
Sample tests shows the most unstable eigenvalues obtained are well separated
from the rest of the spectrum and, further, occurs in conjugate pairs (a more
general result for ideal instabilities which may be shown via consideration of
the adjoint form of the governing equation; see, e.g.,
\citealt{DrazinReid-Stability}).


\subsection{Basic state and full numerical solution}

We consider first the Rayleigh profile as the background flow, with a uniform
background magnetic field. In dimensional form, we take
\begin{equation}\label{s4:profile}
	\Ubar(y)=\begin{cases}\Lambda L, &y>L,\\\Lambda y, &|y|<L,\\
	-\Lambda L, &y<-L,\end{cases}\qquad \Bbar(y)=B_{0}.
\end{equation}
For this problem, $\Jbar=0$ and $\dy\Jbar/\dy y=0$. Scaling by
\begin{equation}
	\tilde{B}=B_{0},\qquad \tilde{T}=\frac{1}{\Lambda},\qquad \tilde{L}=L,
	\qquad \tilde{U}=\Lambda L,
\end{equation}
the resulting non-dimensional parameter is $M=\tilde{B}/\tilde{U}$, a ratio of
the typical Alfv\'en velocity and the shear velocity, effectively a measure of
the field strength, and $\Bbar\rightarrow M$ and $j\rightarrow Mj$ in
\eqref{s4:qj-numerical} (as well as \ref{s4:qeta-numerical}) upon rescaling.
$M\geq1$ corresponds to the regime where there are no linear, normal mode
instabilities. The linear stability properties of this basic state has been
studied by \cite{RayErshkovich83}, who focused on computation of the growth
rates of the instabilities via a shooting method, however not on the structure
of the eigenfunctions.

\begin{figure}
\begin{center}
	\psfrag{0}[cc][cc][0.8]{0}
	\psfrag{0.5}[cc][cc][0.8]{0.5}
	\psfrag{1}[cc][cc][0.8]{1}
	\psfrag{0.25}[cc][cc][0.8]{0.25}
	\psfrag{0.75}[cc][cc][0.8]{0.75}
	\psfrag{0.1}[cc][cc][0.8]{0.1}
	\psfrag{0.2}[cc][cc][0.8]{0.2}
	\psfrag{M}[cc][cc][0.8]{$M$}
	\psfrag{k}[cc][cc][0.8]{$k$}
	\psfrag{full}[cc][cc][0.8]{}
	\includegraphics[width=0.8\textwidth]{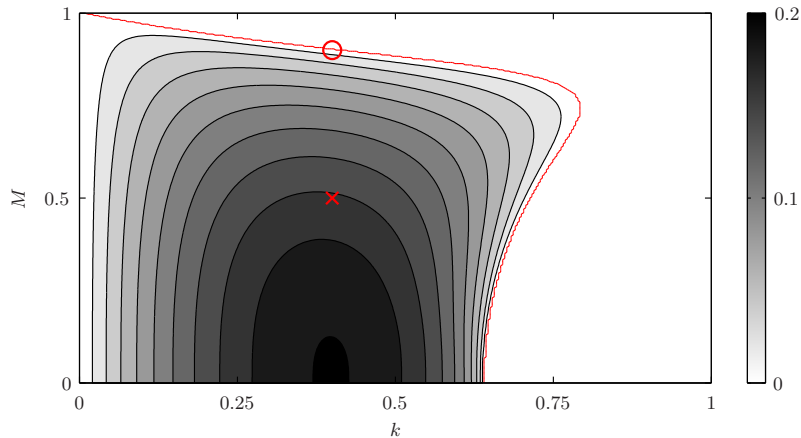}
	\caption{Growth rates $(kc_{i})_{\textnormal{full}}$ of the full numerical
	solution associated with the basic state \eqref{s4:profile}. Here, $c_{r}=0$
	for all unstable modes. The red cross and circle correspond to the
	parameter location associated with the eigenfunction displayed in
	figure~\ref{fig:rayleigh_MHD_eigen}($a,b$) respectively.}
	\label{fig:rayleigh_MHD_full_contour}
\end{center}
\end{figure}

We show in figure~\ref{fig:rayleigh_MHD_full_contour} the growth rates of the
full solution $(kc_{i})_{\textnormal{full}}$ over parameter space; we note that
$c_{r}=0$ for all unstable modes here, a result arising from the symmetries
possessed by the basic state. The calculated growth rates are in agreement with
the results documented in \cite{RayErshkovich83} after taking into account the
different scalings used. The explanation for the shape of the instability region
is that, at $M=0$, there is no phase-locking for short waves since they are too
slow to overcome the background advection. At moderate $M$, these short waves
are now sufficiently fast (from the non-dimensional version of
\ref{s3:eigen-full}) and overcome the background advection, can become
phase-locked, and constructively interfere. At large enough $M$, all waves
become too fast, and phase-locking is not possible.

\begin{figure}
\begin{center}
	\psfrag{0}[cc][cc][0.8]{0}
	\psfrag{0.25}[cc][cc][0.8]{0.25}
	\psfrag{0.5}[cc][cc][0.8]{0.5}
	\psfrag{0.75}[cc][cc][0.8]{0.75}
	\psfrag{1}[cc][cc][0.8]{1}
	\psfrag{1.05}[cc][cc][0.8]{1.05}
	\psfrag{0.95}[cc][cc][0.8]{0.95}
	\psfrag{-0.25}[cc][cc][0.8]{-0.25}
	\psfrag{-0.5}[cc][cc][0.8]{-0.5}
	\psfrag{-0.75}[cc][cc][0.8]{-0.75}
	\psfrag{-0.95}[cc][cc][0.8]{-0.95}
	\psfrag{-1}[cc][cc][0.8]{-1}
	\psfrag{-1.05}[cc][cc][0.8]{-1.05}
	\psfrag{x 10}[cc][cc][0.8]{$\times 10$}
	\psfrag{-3}[ll][ll][0.7]{-3}
	\psfrag{-2}[cc][cc][0.8]{-2}
	\psfrag{2}[cc][cc][0.8]{2}
	\psfrag{-0.002}[cc][cc][0.8]{-0.002}
	\psfrag{0.002}[cc][cc][0.8]{0.002}
	\psfrag{-0.004}[cc][cc][0.8]{-0.004}
	\psfrag{0.004}[cc][cc][0.8]{0.004}
	\psfrag{y}[cc][cc][0.8]{$y$}
	\psfrag{kx/(2pi)}[cc][cc][0.8]{$kx/(2\pi)$}
	\psfrag{M=0.5-k=0.4}[cc][cc][0.8]{($a$)}
	\psfrag{1}[cc][cc][0.8]{1}
	\psfrag{0.9}[cc][cc][0.8]{0.9}
	\psfrag{0.8}[cc][cc][0.8]{0.8}
	\psfrag{-0.8}[cc][cc][0.8]{-0.8}
	\psfrag{-0.9}[cc][cc][0.8]{-0.9}
	\psfrag{0.1}[cc][cc][0.8]{0.1}
	\psfrag{-0.1}[cc][cc][0.8]{-0.1}
	\psfrag{-0.001}[cc][cc][0.8]{-0.001}
	\psfrag{0.001}[cc][cc][0.8]{0.001}
	\psfrag{-0.003}[cc][cc][0.8]{-0.003}
	\psfrag{0.003}[cc][cc][0.8]{0.003}
	\psfrag{M=0.9-k=0.4}[cc][cc][0.8]{($b$)}
	\includegraphics[width=\textwidth]{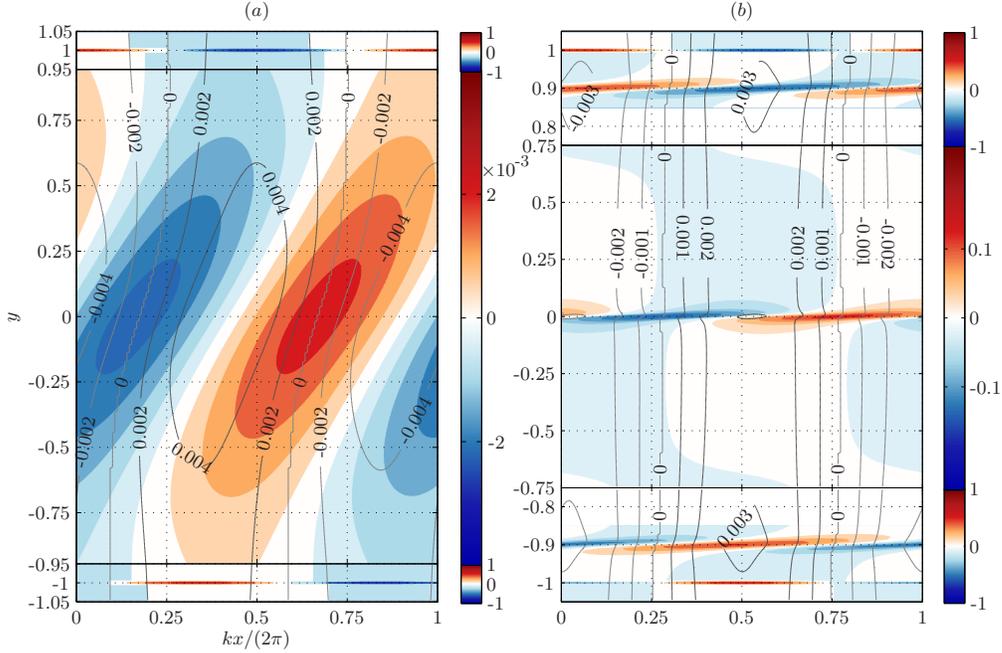}
	\caption{Representative eigenfunctions (normalised by the maximum absolute
	value of $q$) of the profile given in \eqref{s4:profile}, both for $k=0.4$,
	but with ($a$) $M=0.5$, and ($b$) $M=0.9$. The red and blue shading are for
	positive and negative $q$ respectively, and $\eta$ is plotted as labelled
	contours; note the difference in the colour scales used between the
	sub-panels. The $y$-scale is continuous between the panels (note the
	displacement contours are continuous in magnitude) but is not linear for
	display purposes.}
	\label{fig:rayleigh_MHD_eigen}
\end{center}
\end{figure}

We show in figure~\ref{fig:rayleigh_MHD_eigen} two eigenfunctions, one
configuration that is generic for parameter values away from the stability
boundary, and one at the same wavenumber, but at increased $M$ and is a sample
configuration for a case close to marginality; the parameter choices are given
by the red cross and circle in figure~\ref{fig:rayleigh_MHD_full_contour}
respectively.

In figure~\ref{fig:rayleigh_MHD_eigen}($a$), the eigenfunction configuration is
generic in that the dominant contributions to vorticity comes from the two
interfaces at $y=\pm1$, and the contributions in the $|y|\neq1$ region are
comparatively small (note the colour axis scales of the corresponding
sub-panels). The outer two vorticity contributions have a phase-shift that is
close to $\pi/2$ ($0.25$ in the normalised units used in the diagram), and the
middle tilted structure has vorticity contributions that have opposite signs in
between two like-signed vorticity contributions from the interface. The
vorticity eigenfunction is, going from top to bottom, in anti-phase, in
quadrature and in phase with displacement, i.e., $q\sim(-\eta,-\zi\eta,+\eta)$
respectively. The outer two waves are really counter-propagating vorticity
waves, and the inner wave resembles a standing wave. We have further carried out
a decomposition of the vorticity eigenfunction into its constituents using
equation \eqref{s2:q-dim}; in this case, the non-zero terms are $-(\dy\psi/\dy
x)(\dy\Qbar/\dy y)$ and $\Bbar(\dy j/\dy x)$, divided by $\Ubar-c$. At the
interface locations, the vorticity contribution comes from both terms, with the
$-(\dy\psi/\dy x)(\dy\Qbar/\dy y)$ term being the dominant contribution, and the
effect of the $\Bbar(\dy j/\dy x)$ contribution is to alter the magnitude and
the phase-difference between the two waves.

\begin{figure}
\begin{center}
	\psfrag{+qr}[ll][ll][0.8]{$\color{red}+q$}
	\psfrag{-qr}[ll][ll][0.8]{$\color{red}-q$}
	\psfrag{+qb}[ll][ll][0.8]{$\color{blue}+q$}
	\psfrag{-qb}[ll][ll][0.8]{$\color{blue}-q$}
	\psfrag{+qg}[ll][ll][0.8]{$\color{dark-green}+q$}
	\psfrag{-qg}[ll][ll][0.8]{$\color{dark-green}-q$}
	\psfrag{top}[ll][ll][0.8]{$q\sim-\eta$}
	\psfrag{middle}[ll][ll][0.8]{$q\sim-\zi\eta$}
	\psfrag{bottom}[ll][ll][0.8]{$q\sim+\eta$}
	\includegraphics[width=\textwidth]{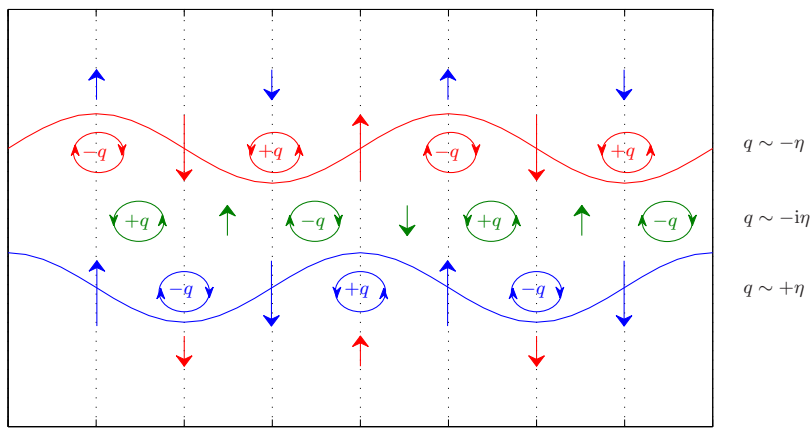}
	\caption{Schematic of how the critical layer contribution plays a role in
	the dynamics. Without the centre contribution denoted by the green parts,
	the dynamics of the two waves ($q\sim-\eta$ and $q\sim+\eta$) are as in
	figure~\ref{fig:CRW_basic}. The centre wave acts as a standing wave with
	$q\sim-\zi\eta$, with a resulting vorticity configuration as depicted in the
	figure. Then it is seen that the extra contributions acts against the
	vorticity anomalies associated with the counter-propagating waves, and is
	thus a stabilising effect.}
	\label{fig:CRW_crit}
\end{center}
\end{figure}

Figure~\ref{fig:rayleigh_MHD_eigen}($a$) may roughly be represented
schematically as figure~\ref{fig:CRW_crit}. With the configuration depicted, we
see that the presence of the middle standing wave counteracts the effects of the
vorticity anomalies associated with the outer counter-propagating waves. If we
were to make the interface assumption, we neglect the contributions away from
the interfaces and remove this centre contribution, and we thus expect to
over-estimate (i), the growth rates, and (ii), the propagation speed, the region
where phase-locking is possible and thus the size of the instability region. We
would expect the over-estimation to be most significant when the field strength
is large, and for short waves. We expect this for short waves because the
interaction decreases exponentially with wavenumber, so the overall constructive
interference between the counter-propagating waves is weaker for short waves
than long waves, and it is short waves that are too slow to overcome the
background advection.

The phase relation for the middle contribution $q\sim-\zi\eta$ may be obtained
from considering the vorticity equation \eqref{s2:q-dim}. Away from $y=\pm1$, we
have, for this problem,
\begin{equation}
	q=\frac{M^{2}j}{(\Ubar-c_{r})-\zi c_{i}}.
\end{equation}
We may suppose that $q$ is likely to be maximised at the location where
$\Ubar-c_{r}=0$, which is $y=0$ here since $c_{r}=0$, resulting in the relation
$q\sim \zi M^{2}j/c_{i}$. From relation \eqref{s2:a-eta-relation}, we have
$j\sim\grad^{2}\eta$, and so $q\sim-\zi M^{2}\eta/c_{i}$, as required. This is
analogous to what was found in the stratified setting for
\cite{Rabinovich-et-al11}. The effect of the critical level is expected to be
more pronounced when we are near marginality.

As we approach marginality via increasing $k$ and/or increasing $M$, the tilting
of the middle structure increases, becomes thinner in the cross-stream extent,
and eventually splits into three tilted structures. A sample case where we
approach marginality by increasing $M$ is shown in
figure~\ref{fig:rayleigh_MHD_eigen}($b$). The first thing to note is that
vorticity generation is not just significant at the location where
$\Ubar(y)-c_{r}=0$, but also at around $y=\pm M$. The mechanistic interpretation
for instability and stabilisation however is not significantly altered. There is
perhaps some cancellation of the contributions due to the tilted structures,
but, schematically, we still have two counter-propagating waves with a standing
wave in between that counteracts the vorticity contributions associated with the
counter-propagating waves, again as in figure~\ref{fig:CRW_crit}.

The structures at $y=\pm M$ are perhaps not too surprising once we note that
Alfv\'en waves are non-dispersive and have non-dimensional phase speed $\cA=\pm
M$ in our setting, regardless of the wave number. Thus, from a physical point of
view, we have forced, short wavelength standing Alfv\'en waves since
$\Ubar(y)-\cA=0$ at these locations. From a mathematical point of view, we
recall that, equivalently, the modal problem with general basic state in
two-dimensional incompressible MHD in non-dimensional form is governed by the
second-order differential equation
\begin{equation}
	\frac{\mathrm{d}}{\mathrm{d}y}
	\left(S^{2}(y)\frac{\mathrm{d}\eta}{\mathrm{d}y}\right)
	-k^{2}S^{2}(y)\eta=0,\qquad
	S^{2}(y)=(\Ubar(y)-c)^{2}-M^{2}\Bbar^{2}(y).
\end{equation}
If we take the view that critical levels are where the governing eigenvalue
differential equations break down, the critical levels can occur when
If we take the view that critical levels are where the governing eigenvalue
differential equations break down, the critical levels can occur when
$\Ubar(y)-c=0$ and also where $\Ubar(y)-c=\pm M\Bbar(y)$, with the latter
associated with Alfv\'en waves. The appearance of multiple critical levels is
not an isolated feature in MHD, and appears for example in shear flow problems
on the $f$-plane \citep[e.g., appendix B of][]{Lott03}.

The second point to note for figure~\ref{fig:rayleigh_MHD_eigen}($b$) is the
change in the phase-shift of the counter-propagating components compared to
figure~\ref{fig:rayleigh_MHD_eigen}($a$), even though the same wavenumber was
chosen. The stabilisation does not solely come from increasing the strength of
the centre contribution, unlike in one of the examples considered by
\cite{Rabinovich-et-al11}. By changing the field strength, we change both the
strength of the critical layer contribution and the wave properties at the
interfaces. Although the critical layer contribution is now stronger, the
phase-shift between the vorticity contributions of the outer waves are also
approaching an anti-phase configuration, and both effects contribute to the
reduction in growth rate. It is also interesting to recall that, when the two
outer vorticity contributions are in anti-phase, this implies that the
corresponding displacement is in phase \cite[e.g.,][]{Heifetz-et-al04a}, as seen
in the $\eta$ eigenfunction. In other words, the whole region is undulating as
one; this is physically consistent in that, as we increase the field strength,
the magnetic field imparts more `stiffness' to the fluid, and the fluid is
forced to undulate as a whole.

As a final point, figure~\ref{fig:rayleigh_MHD_eigen}($b$) is generic for
eigenfunctions near marginal stability in the sense that, as $c_{i}$ goes to
zero, the tilted structures become thinner and the configuration goes to one
where there is no longer constructive interference due to the vorticity
associated with the counter-propagating modes being in phase or in anti-phase.
We have shown here the case where vorticity becomes increasingly out of phase,
but of course they may also become increasingly in phase depending on where we
are on the stability boundary. There is an intermediate region where the
stabilisation is solely due to the strengthening of the centre contribution, but
this is a somewhat special case and, generically, both the changes to
phase-shift and strengthening of the critical layer contributes to the
neutralisation of the instability.


\subsection{Interfacial wave dynamics}\label{s4-analytical}

From figure~\ref{fig:rayleigh_MHD_eigen}($a$), we observe that the contribution
in the centre can be small, so we may suppose that taking solutions of the form
\begin{equation}\label{s4:solu-form}
	q=\qhat_{1}\delta(y-1)+\qhat_{2}\delta(y+1),\qquad
	j=\jhat_{1}\delta(y-1)+\jhat_{2}\delta(y+1)
\end{equation}
has the potential to be a reasonable approximation to the full solution, at
least away from marginality. Since this neglects the standing wave contribution
that counteracts the two counter-propagating waves, we expect the resulting
solutions have, compared to the solutions presented in
figure~\ref{fig:rayleigh_MHD_full_contour}, larger growth rates and a larger
region of instability.

We start from the ($q,\eta$) formulation with equation \eqref{s2:q-dim} and
\eqref{s2:eta-dim}. From the inversion relation \eqref{s2:inversion}, we have,
in non-dimensional form,
\begin{equation}\label{s4:inversion-relation}
	A_{1,2}=-\frac{1}{2k}(\jhat_{1,2}+\jhat_{2,1}\ex^{-2k}),\qquad
	\psi_{1,2}=-\frac{1}{2k}(\qhat_{1,2}+\qhat_{2,1}\ex^{-2k}).
\end{equation}
Our vorticity equation \eqref{s2:q-dim} will have $\jhat$ involved so we also
need expressions for $\jhat_{1,2}$. Using $A=\Bbar\eta$ from
\eqref{s2:a-eta-relation}, we obtain from equation \eqref{s4:inversion-relation}
\begin{equation}\label{s4:j-relation}
	\jhat_{1,2}=-\frac{2k}{1-\ex^{-4k}}
	(\eta_{1,2}-\eta_{2,1}\ex^{-2k}).
\end{equation}
Substituting accordingly, the governing equations \eqref{s2:q-dim} and
\eqref{s2:eta-dim} becomes
\begin{subequations}\label{s4:nondim}\begin{align}
	\left(\frac{\dy}{\dy t}\pm\zi k\right)\qhat_{1,2}&=
	\pm\frac{\zi}{2}(\qhat_{1,2}+\qhat_{2,1}\ex^{-2k})
	-\frac{2kM^{2}}{1-\ex^{-4k}}(\eta_{1,2}-\eta_{2,1}\ex^{-2k}),
	 \label{s4:q-nondim}\\
	\left(\frac{\dy}{\dy t}\pm\zi k\right)\eta_{1,2}&=
	-\frac{\zi}{2}(\qhat_{1,2}+\qhat_{2,1}\ex^{-2k})
	\label{s4:eta-nondim}.
\end{align}\end{subequations}
In writing equations \eqref{s4:nondim}, the fundamental assumption is that the
other interface exists, which results in the appearance of exponential factors
$(1-\ex^{-4k})^{-1}$ multiplying some of the $\eta$ terms. This is unlike the
stratified case considered in \cite{Rabinovich-et-al11}. Although the
displacement is related to the perturbation magnetic potential and buoyancy for
the respective cases, in the MHD case the current also appears in the vorticity
equation, while there is no analogue of this in the stratified case.

Considering modal solutions, the system \eqref{s4:nondim} has closed form
solutions given by
\begin{equation}\label{s4:c_solution}
	c=\pm\sqrt{1-\frac{1}{2k}+\frac{1-\ex^{-4k}}{8k^{2}}+M^{2}
	\pm\sqrt{\frac{1}{4k^{2}}\left(1-\frac{1-\ex^{-4k}}{4k}\right)^{2}
	+2M^{2}\xi}},
\end{equation}
where
\begin{equation}
	\xi=1-\frac{1}{k}+\frac{1-\ex^{-4k}}{8k^{2}}
	+\frac{1+\ex^{-4k}}{1-\ex^{-4k}}.
\end{equation}
When $M=0$, \eqref{s4:c_solution} reduces to the hydrodynamic solutions
$c=\pm\sqrt{(1-2k)^{2}-\ex^{-4k}}/2k$ \citep[e.g.,][]{DrazinReid-Stability}. For
$k\ll1$, a straight forward asymptotic analysis of \eqref{s4:c_solution} yields
$c=\zi\sqrt{1-M^{2}}$, which is the vortex sheet result in incompressible MHD
\citep[e.g.,][]{Chandrasekhar-Stability}.

\begin{figure}
\begin{center}
	\psfrag{0}[cc][cc][0.8]{0}
	\psfrag{0.25}[cc][cc][0.8]{0.25}
	\psfrag{0.5}[cc][cc][0.8]{0.5}
	\psfrag{0.75}[cc][cc][0.8]{0.75}
	\psfrag{1}[cc][cc][0.8]{1}
	\psfrag{0.1}[cc][cc][0.8]{0.1}
	\psfrag{0.2}[cc][cc][0.8]{0.2}
	\psfrag{M}[cc][cc][0.8]{$M$}
	\psfrag{k}[cc][cc][0.8]{$k$}
	\psfrag{int}[cc][cc][0.8]{($a$)}
	\psfrag{diff}[cc][cc][0.8]{($b$)}
	\includegraphics[width=\textwidth]{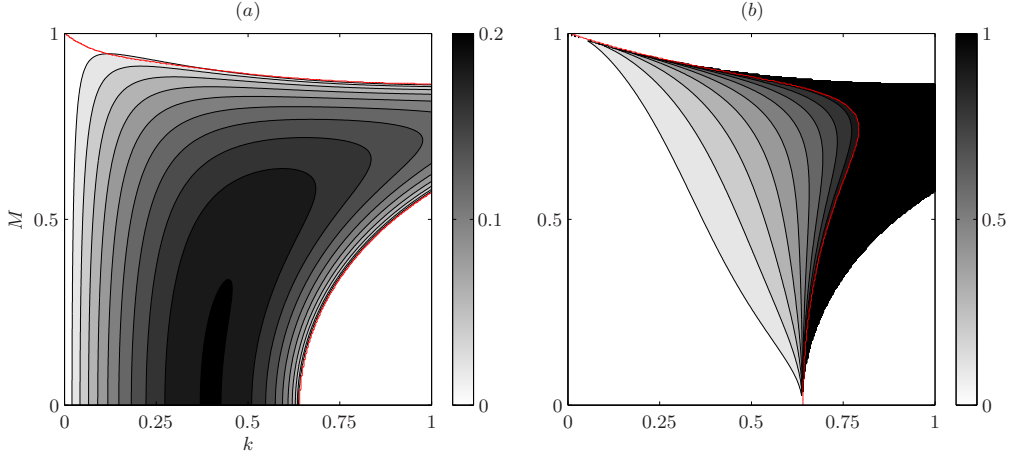}
	\caption{Results from taking only two interfaces, for the profile given in
	\eqref{s4:profile}. ($a$) shows the growth rates
	$(kc_{i})_{\textnormal{int}}$ of the approximated solution from
	\eqref{s4:c_solution}. ($b$) shows the weighted error
	$E=1-(kc_{i})_{\textnormal{full}}/(kc_{i})_{\textnormal{int}}$; the red line
	here is the stability boundary of the full solution in
	figure~\ref{fig:rayleigh_MHD_full_contour}.}
	\label{fig:rayleigh_MHD_contour_E}
\end{center}
\end{figure}

Figure~\ref{fig:rayleigh_MHD_contour_E}($a$) shows the growth rates
$(kc_{i})_{\textnormal{int}}$ of the unstable branch of \eqref{s4:c_solution}.
Like the full solution, the unstable modes here have $c_{r}=0$. In
figure~\ref{fig:rayleigh_MHD_contour_E}($b$) we show contours of the weighted
error $E=1-(kc_{i})_{\textnormal{full}}/(kc_{i})_{\textnormal{int}}$; when
$E=0$, the approximated solution \eqref{s4:c_solution} completely agrees with
the full solution, whilst $E=1$ shows the approximated solution predicts
instability when it is otherwise absent in the full solution. In agreement with
the hypotheses, we over-estimate growth rates as well as regions of instability,
with the over-estimation being most significant in the short wave regime and
near the stability boundary.

\begin{figure}
\begin{center}
	\psfrag{0}[cc][cc][0.8]{0}
	\psfrag{1}[cc][cc][0.8]{1}
	\psfrag{2}[cc][cc][0.8]{2}
	\psfrag{0.2}[cc][cc][0.8]{0.2}
	\psfrag{-0.2}[cc][cc][0.8]{-0.2}
	\psfrag{0.3}[cc][cc][0.8]{0.3}
	\psfrag{-0.3}[cc][cc][0.8]{-0.3}
	\psfrag{kci}[cc][cc][0.8]{$kc_{i}$}
	\psfrag{cr}[cc][cc][0.8]{$c_{r}$}
	\psfrag{k}[cc][cc][0.8]{$k$}
	\psfrag{M=0.5}[cc][cc][0.8]{($a$)}
	\psfrag{M=0.75}[cc][cc][0.8]{($b$)}
	\psfrag{M=0.9}[cc][cc][0.8]{($c$)}
	\includegraphics[width=\textwidth]{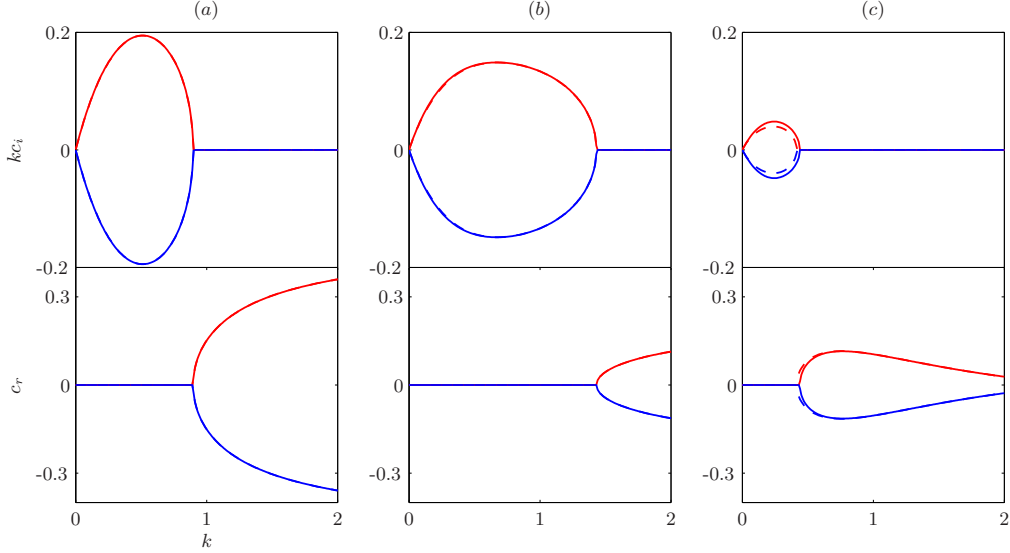}
	\caption{Line plots of the growth rate $kc_{i}$ (top row) and the phase
	speed $c_{r}$ (bottom row) associated with the growing and decaying
	solutions of the full solution \eqref{s4:c_solution}, denoted by solid
	lines, and the $\chi=0$ solutions \eqref{s4:chi0-solution}, denoted by
	dashed lines. These are evaluated at fixed $M$, with ($a$) $M=0.5$,
	($b$) $M=0.75$, and ($c$) $M=0.9$. The $\chi=0$ solution is, for the most
	part, indistinguishable from the analytical solution.}
	\label{fig:rayleigh_MHD_line}
\end{center}
\end{figure}

Previous authors analysing the stratified problem \citep{Harnik-et-al08,
Rabinovich-et-al11} have observed that, since there are two wave branches, the
modal solutions will have a pro- and counter-propagating component. The
existence of the two components are due to the fact that there is an
intermediate step linking displacement with vorticity. In our schematic
presented above in figure~\ref{fig:CRW_crit}, the instability results from the
interaction of counter-propagating modes, so it is informative to investigate
the role of the pro-propagating mode. A similar analysis to
\cite{Rabinovich-et-al11} may be carried out here, by taking into account the
asymmetric non-dimensional eigenstructure given by
\begin{equation}\label{s4:eigen-rossby-nondim}
	\qhat^{\pm}_{1}=2k\cR^{\pm}\eta^{\pm}_{1},\qquad 
	\qhat^{\pm}_{2}=-2k\cR^{\mp}\eta^{\pm}_{2},\qquad
	\cR^{\pm}=-\frac{1}{4k}
	\pm\sqrt{\left(\frac{1}{4k}\right)^{2}+M^{2}}.
\end{equation}
We may transform the system of equations \eqref{s4:nondim} into a system of
equations in terms of ($\eta_{1}^{\pm},\eta_{2}^{\pm}$) via a direct
substitution, a self-similarity transform \citep[e.g.,][]{Harnik-et-al08}, or
otherwise. We come to essentially the same conclusions as
\cite{Rabinovich-et-al11}, where, when there is an instability, the
pro-propagating mode should satisfy the relation
$(\eta_{1}^{+},\eta_{2}^{-})=-\chi(\eta_{2}^{+},\eta_{1}^{-})$, where
$\chi\in[0,1]$, i.e., the pro-propagating mode on one flank is smaller by a
factor of $\chi$ and in anti-phase with the counter-propagating mode on the
other flank. The existence of the pro-propagating mode is to provide extra
hindering to the counter-propagating modes, with its effect being most
significant in the $k\ll1$ regime. One could consider taking $\chi=0$
approximation that artificially removes the pro-propagating mode; the resulting
analytical solution is
\begin{equation}\label{s4:chi0-solution}
	c=\pm\sqrt{\left(1+\cR^{-}+
	\frac{M^{2}\ex^{-4k}}{(1-\ex^{-4k})(\cR^{+}-\cR^{-})}\right)^{2}
	-\left(\cR^{-}-\frac{2kM^{2}}{1-\ex^{-4k}}\right)^{2}
	\left(\frac{\ex^{-2k}}{\cR^{+}-\cR^{-}}\right)^{2}}.
\end{equation}
We show in figure~\ref{fig:rayleigh_MHD_line} several line plots of the
analytical solution \eqref{s4:c_solution} and the $\chi=0$ solution
\eqref{s4:chi0-solution}. The differences between the solutions in this case are
so slight that they are only distinguishable at high values of $M$. This points
to the scenario that, although the pro-propagating mode must exist as part of
the physics, its effect on the instability for this profile is almost negligible
compared to the counter-propagating mode. One may obtain the numerical values of
$\chi$ by computing the eigenfunctions (which are just complex numbers in this
case), and taking $\chi = |\eta_1^+|/|\eta_2^+| = |\eta_2^-|/|\eta_1^-|$;
although not shown here, we have $\chi\lesssim0.1$ over most of the region where
there is instability.

\section{Stable profile destabilised by a spatially varying magnetic field}

We now carry out a similar investigation for the case where a stable flow is
destabilised by a spatially varying magnetic field. We consider the dimensional
basic state
\begin{equation}\label{s5:profile}
	\Ubar(y)=\Lambda y,\qquad
	\Bbar(y)=\begin{cases}+\Gamma y, &y>L,\\
	+\Gamma L, &|y|<L,\\ -\Gamma y, &y<-L,
	\end{cases}
\end{equation}
and $\dy\Qbar/\dy y=0$ here. Scaling by
\begin{equation}
	\tilde{B}=\Gamma L,\qquad \tilde{T}=\frac{1}{\Lambda},\qquad \tilde{L}=L
	\qquad \tilde{U}=\Lambda L,
\end{equation}
the non-dimensional parameter is again $M=\tilde{B}/\tilde{U}$, and
$(\Bbar,\Jbar,j)\rightarrow M(\Bbar,\Jbar,j)$ after rescaling. This
piecewise-linear magnetic field profile (resembling a wake) is chosen so that it
satisfies $\Bbar\Delta\Jbar<0$. This profile choice is partly inspired by the
parabolic magnetic field profiles $\Bbar\sim y^{2}$ considered by
\cite{ChenMorrison91} and \cite{TatsunoDorland06} and the single interface
profile considered by \cite{Stern63}. One notable difference however is that
this profile has a well-defined region of maximum $|\dy\Jbar/\dy y|$, whilst the
profiles considered previously generally have $\dy\Jbar/\dy
y=\textnormal{const}$ throughout the domain, so this profile is not intended to
be directly comparable to those previous studies. This case also has some
similarities to the problem where there is a linear shear with two density jumps
(the Taylor--Caulfield type instability), considered previously by, for example,
\cite{Rabinovich-et-al11}; we have here instead two jumps in the current
profile. Again, $M=1$ is the cutoff for linear, normal mode instability.

\begin{figure}
\begin{center}
	\psfrag{0}[cc][cc][0.8]{0}
	\psfrag{0.5}[cc][cc][0.8]{0.5}
	\psfrag{1}[cc][cc][0.8]{1}
	\psfrag{0.25}[cc][cc][0.8]{0.25}
	\psfrag{0.75}[cc][cc][0.8]{0.75}
	\psfrag{0.03}[cc][cc][0.8]{0.03}
	\psfrag{0.06}[cc][cc][0.8]{0.06}
	\psfrag{M}[cc][cc][0.8]{$M$}
	\psfrag{k}[cc][cc][0.8]{$k$}
	\psfrag{full}[cc][cc][0.8]{}
	\includegraphics[width=0.8\textwidth]{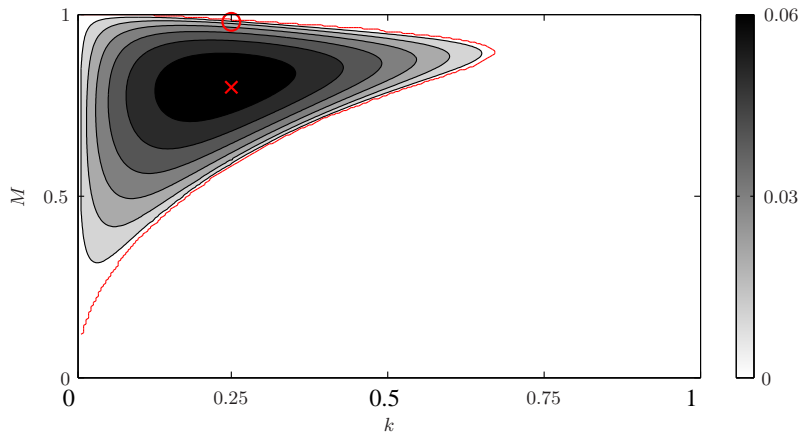}
	\caption{Growth rates $(kc_{i})_{\textnormal{full}}$ of the full numerical
	solution associated with the basic state \eqref{s5:profile}. Here, $c_{r}=0$
	for all unstable modes. The red cross and circle corresponds to the
	parameter location associated with the eigenfunction displayed in
	figure~\ref{fig:couette_MHD_eigen}($a,b$) respectively.}
	\label{fig:couette_MHD_full_contour}
\end{center}
\end{figure}

Figure~\ref{fig:couette_MHD_full_contour} shows contours of
$(kc_{i})_{\textnormal{full}}$ for this instability problem, and, again,
$c_{r}=0$ for the unstable modes. The shape of the instability region in
parameter space may again be appropriately justified. At small $M$, waves are
too slow to overcome the background advection except at $k\ll1$. As $M$
increases, waves become sufficiently fast, overcome the background advection and
achieve phase-locking. For large enough $M$, all waves are too fast to
phase-lock.

\begin{figure}
\begin{center}
	\psfrag{0}[cc][cc][0.8]{0}
	\psfrag{0.25}[cc][cc][0.8]{0.25}
	\psfrag{0.5}[cc][cc][0.8]{0.5}
	\psfrag{0.75}[cc][cc][0.8]{0.75}
	\psfrag{1}[cc][cc][0.8]{1}
	\psfrag{0.95}[cc][cc][0.8]{0.95}
	\psfrag{-0.25}[cc][cc][0.8]{-0.25}
	\psfrag{-0.5}[cc][cc][0.8]{-0.5}
	\psfrag{-0.75}[cc][cc][0.8]{-0.75}
	\psfrag{-0.95}[cc][cc][0.8]{-0.95}
	\psfrag{-1}[cc][cc][0.8]{-1}
	\psfrag{1}[cc][cc][0.8]{1}
	\psfrag{-1}[cc][cc][0.8]{-1}
	\psfrag{0.125}[cc][cc][0.8]{0.125}
	\psfrag{-0.125}[cc][cc][0.8]{-0.125}
	\psfrag{0.25}[cc][cc][0.8]{0.25}
	\psfrag{-0.25}[cc][cc][0.8]{-0.25}
	\psfrag{0.03}[cc][cc][0.8]{0.03}
	\psfrag{-0.03}[cc][cc][0.8]{-0.03}
	\psfrag{0.015}[cc][cc][0.8]{0.015}
	\psfrag{-0.015}[cc][cc][0.8]{-0.015}
	\psfrag{y}[cc][cc][0.8]{$y$}
	\psfrag{kx/(2pi)}[cc][cc][0.8]{$kx/(2\pi)$}
	\psfrag{M=0.8-k=0.25}[cc][cc][0.8]{($a$)}
	\psfrag{M=0.98-k=0.25}[cc][cc][0.8]{($b$)}
	\includegraphics[width=\textwidth]{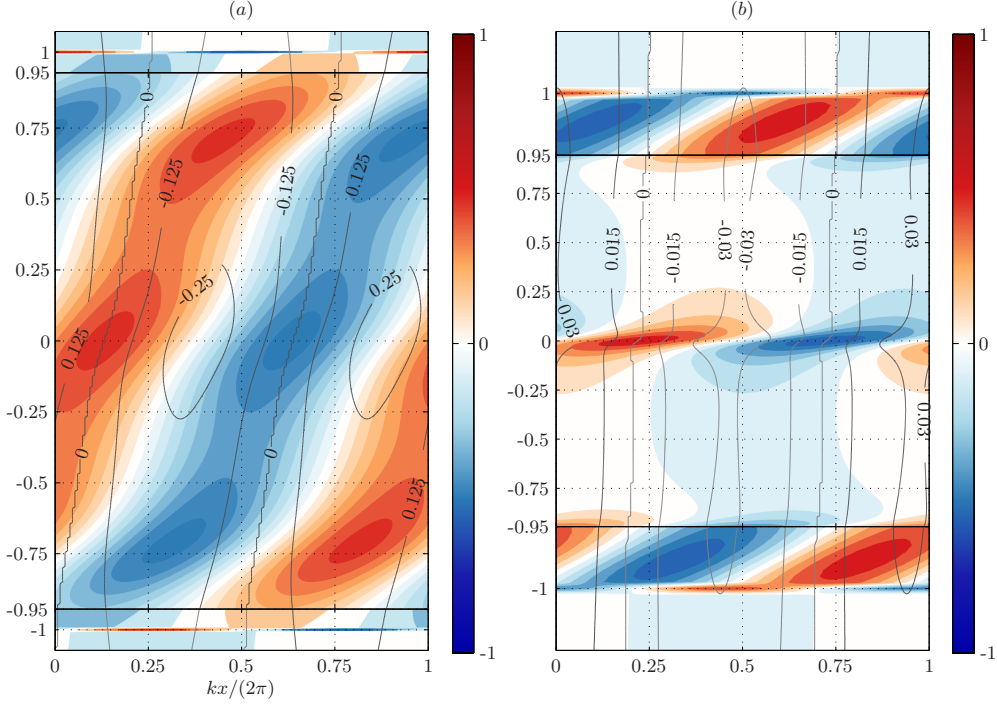}
	\caption{Representative eigenfunctions (normalised by the maximum absolute
	value of the vorticity) of the profile given in \eqref{s5:profile}, both for
	$k=0.25$, but with ($a$) $M=0.8$, and ($b$) $M=0.98$. The red and blue
	shading are for positive and negative vorticity respectively, and the
	displacement is plotted as labelled contours. The $y$-scale is continuous
	between the panels (note the displacement contours are continuous in
	magnitude) but is not linear for display purposes.}
	\label{fig:couette_MHD_eigen}
\end{center}
\end{figure}

In figure~\ref{fig:couette_MHD_eigen} we show two eigenfunctions, one that is
representative of a growing mode away from the stability boundary, and one for a
case close to marginality; again, the parameter locations are given respectively
by the red cross and circle in figure~\ref{fig:couette_MHD_full_contour}. For
both cases, we notice that the same contour levels are used for all three
sections, unlike the previous instability problem displayed in
figure~\ref{fig:rayleigh_MHD_eigen}. This immediately suggests that considering
interfacial wave solutions only at $y=\pm1$ as in \eqref{s4:solu-form} is going
to be an overly drastic approximation, since we will be neglecting contributions
that are comparable in size to those at the interface.

Like the previous case, we have significant structures appearing at $y=\pm M$,
the locations where we have forced, stationary Alfv\'en waves. What is different
in this case is that the schematic presented in figure~\ref{fig:CRW_crit}
applies for the inner tilted structures, and so it is Alfv\'en waves at $y=\pm
M$ rather than the waves supported by the interfaces that drive the instability.
There are several extra interactions between the structures (affecting wave
propagation and interaction) that lead to the overall instability eigenfunction
in figure~\ref{fig:couette_MHD_eigen}. Comparing between
figure~\ref{fig:couette_MHD_eigen}($a,b$), we observe again that it is a mix of
changes to the overall phase-shift between the structures that lead to the
neutralisation of the instability. Notice also that, since we are approaching
marginality by increasing $M$ in figure~\ref{fig:couette_MHD_eigen}($b$), the
displacement contours become increasingly in phase, as in the previous example
in figure~\ref{fig:rayleigh_MHD_eigen}($b$).

We considered decomposing the full vorticity eigenfunction into its constituents
using \eqref{s2:q-dim}; in this case, it is only the $\Bbar(\dy j/\dy x)$ and
the $(\dy A/\dy x)(\dy\Jbar/\dy y)$, divided by $\Ubar-c$, that contributes to
the overall vorticity. Like the problem considered in \S4, all the contributions
away from $y=\pm 1$ come solely from the former term, whilst both contribute to
the vorticity eigenfunction at the interfaces. It may be seen that while the
latter term is associated with a particular contribution, the former term has
associated a contribution that is in anti-phase and marginally larger than this,
thus resulting in what is seen in figure~\ref{fig:couette_MHD_eigen}. It is
interesting in that although it looks like the contributions at $y=\pm1$ plays a
relatively minor role in the eigenfunction figure~\ref{fig:couette_MHD_eigen} as
the cause for instability, it really must exist for the overall interaction
resulting in the instability, via hindering/helping of the wave propagation over
the domain.

We should stress here that the instability in figure~\ref{fig:couette_MHD_eigen}
really does require the presence of two current jumps to operate, while the
$\Bbar(\dy j/\dy x)$ term acts to modify the resulting interaction. To support
this claim, we have carried out calculations where we only have one jump. In
this setting, there is no instability since the contributions on the other
interface, its adjacent tilted structure and the standing wave in the centre are
absent. The resulting vorticity eigenfunction has the interface and adjacent
tilted structure in anti-phase, and there is no constructive interference there.
Removing both the jumps (i.e., linear shear flow with a uniform background
magnetic field) also does not result in instability. For the profile
\eqref{s5:profile} we consider here, the strength of the current gradient and
background magnetic field are simultaneously controlled by $M$. We considered a
modified problem where the two parameters may be varied independently; we also
arrive at similar conclusions to the one presented here.

We may take a similar approach to the work detailed in \S\ref{s4-analytical}, by
neglecting contributions away from the interfaces, which results in analytical
solutions that may be analysed accordingly. However, we do not expect this to
provide an accurate approximation for this particular choice of basic state,
since we are neglecting contributions that are of the same order of magnitude as
the ones at the interface. A comparison of the full numerical solutions with the
resulting analytical solution shows that the analytical solution grossly
over-estimates growth rates and the region of instability. In light of the poor
comparison of the analytical result with the correct full numerical solution, we
omit here the results that are counterparts to those presented in
\S\ref{s4-analytical}.


\section{Conclusion and discussion}

We have extended the interacting vorticity wave formalism to the MHD setting to
provide a physical interpretation of the instability mechanism for MHD shear
instabilities. In this framework, the existence of instability depends on
whether the choice of basic state allows vorticity waves to resonate, and
whether the field stabilises or destabilises is dependent on how the resulting
configuration affects wave properties. We have demonstrated that vorticity
generation occurs at the locations where the background magnetic field and the
background shear are both non-zero so, unlike certain hydrodynamic cases,
critical layers must play a role in the dynamics even for piecewise-linear basic
states.

To demonstrate the modifications to the underlying instability mechanism by MHD
effects, and to compare with previous results and to evaluate the limitations of
the interfacial wave assumption, we considered the instability characteristics
of two piecewise-linear basic states, one where the field stabilises the
unstable flow, and the other where the field destabilises the stable flow. The
first example considered is the Rayleigh profile. The growth rate contours of
the full solution agrees with the previous results of \cite{RayErshkovich83};
our contribution here is to rationalise the shape of the instability region via
the properties of the phase-speeds of the wave propagation. Plots of the
eigenfunctions shows that we effectively have a standing wave structure at
$\Ubar(y)-c=0$ in between two counter-propagating modes, and may be
schematically represented by figure~\ref{fig:CRW_crit}. Additionally, we are
able to predict the phase relations of this standing wave contribution from the
equations. As we approach marginality, one interesting feature is that we have
additional structures appearing at the levels where $\Ubar(y)-c=\pm M\Bbar(y)$,
and these are critical levels that correspond to locations where we have forced,
stationary Alfv\'en waves. Changing the field strength affects both the strength
of the critical layer and phase-shifts of the waves, and although there will be
special cases where marginality is achieved when the critical layer contribution
overwhelms the other contributions, generically speaking, it is a combination of
the two effects that leads to neutralisation of the instability. In this
example, we argued that the dynamics of two interfacial waves can serve as a
reasonable approximation to the full problem. We predicted and found that such
an approximation over-estimates the growth rate and the region of instability.
Appropriate analyses in the manner of \cite{Rabinovich-et-al11} were performed
to explore the instability characteristics under the interfacial wave
assumption.

The second example we considered is a linear shear flow destabilised by a
spatially varying background magnetic field. The magnetic field profile was
inspired by the parabolic profile $\Bbar(y)\sim y^{2}$ considered by both
\cite{ChenMorrison91} and \cite{TatsunoDorland06}, although we stress that the
results are not entirely comparable since $\dy\Jbar/\dy y$ is constant
throughout the domain for the parabolic profiles. With regards to the
eigenfunction, a robust feature is that, like the previous example, tilted
structures with significant contributions of vorticity exists at the critical
levels. The schematic in figure~\ref{fig:CRW_crit} however applies instead to
the forced, stationary Alfv\'en waves located at $\Ubar(y)-c=\pm M\Bbar(y)$ and
is counteracted by the standing wave contribution associated with the critical
level $\Ubar(y)-c=0$. The principal interaction driving the instability are from
these stationary Alfv\'en waves away from the interfaces rather than from
interfacial waves.

Part of the reason for employing piecewise-linear profiles is to obtain a
simplified problem for understanding the dynamics leading to instability, as
well as for comparison with previous works on a similar topic. One important
point we highlighted is that one needs to be careful when making the interfacial
wave assumption, since vorticity generation is generally not localised. The
non-local generation of vorticity occurs more generally when considering the
instability problem for smooth basic states (arguably more realistic for
modelling purposes), and the resulting eigenstructure generically has a spatial
dependence on the cross-stream coordinate. Although waves are then not as
well-defined, one may wonder whether the same mechanistic interpretation
summarised by figure~\ref{fig:CRW_crit} here is schematically correct. One study
that supports this was reported in one of the authors' PhD thesis
\citep{Mak-thesis}. For calculations of the profile $\Ubar(y)=\tanh(y)$ and a
uniform background magnetic field, plots of the eigenfunction showing structures
similar to figure~\ref{fig:rayleigh_MHD_eigen}($a$) here were found. We expect
analogous structures to appear in eigenfunctions from calculations with other
smooth basic states, demonstrating that shear instabilities may be interpreted
as the mutual interaction of vorticity regions. One other possible scenario in
smooth profiles is that the overall eigenfunction could look schematically like
figure~\ref{fig:CRW_eigenstructure2}($b$), occurring for example when the basic
state gradients are weak/non-existent, although this has not been found for the
examples considered here.

Although we have focussed on modal instabilities here, the formulation is kept
in the $\dy{\boldsymbol{\eta}}/\dy t =\boldsymbol{\mathsf{A}\eta}$ form (where
$\boldsymbol{\eta}$ denotes the state vector and $\boldsymbol{\mathsf{A}}$
denotes the appropriate operator) so that it may also be used to investigate
non-normal mode instabilities and transient growth
\citep[e.g.,][]{ConstantinouIoannou11, GuhaLawrence14}.

Beyond incompressible MHD, this wave interaction framework, together with the
previous works on for the stratified case \citep{Harnik-et-al08,
Rabinovich-et-al11, Carpenter-et-al10, Carpenter-et-al13, GuhaLawrence14}
may perhaps explain the observations made in the previous work of
\cite{Leconanet-et-al10}, where they consider the shear instability problem in a
stratified fluid, with a background magnetic field. The field in that case can
stabilise or destabilise, and we suspect this is most likely due to whether the
vorticity wave modes supported by the choice of the basic state can interact
accordingly, leading to instability; we suspect this is why the Richardson
number or Miles--Howard criterion \citep[e.g.,][]{Miles61} is not necessarily
applicable to stratified MHD shear flows.

Our profiles were chosen so that, when $M\geq1$, $|B|\geq|U|$
pointwise everywhere, and a stability theorem forbids normal mode
instabilities. Profiles violating this condition locally do suffer
instabilities, and, in particular, have been found for studies in both
two-dimensional, spherical, incompressible and shallow water MHD \citep[see
the recent review by][]{GilmanCally07}. Although this particular scenario is
not one we have addressed here, a extension of our interpretation to spherical
MHD appears possible \cite[a hydrodynamic extension in spherical co-ordinates
was given by][]{Methven-et-al05c}. These instabilities may be explained by
energetic arguments, and an extension of our mechanistic interpretation will
serve to complement the existing explanation. A formulation of the problem
in spherical MHD allows for a more appropriate comparison to the existing
results, and this is currently under investigation by the authors.

This work was initiated whilst JM and EH were visiting MISU, with EH supported
by the Rossby visiting fellowship of the International Meteorological Institute
of Sweden. EH is grateful to Michael Mond for fruitful discussion and to his
teacher Alexander Ershkovich for his insight and wisdom. The order of authorship
is alphabetical.

\appendix


\section{Some formal analogies with the stratified problem}\label{appendix1}

Using \eqref{s2:a-eta-relation} to rewrite \eqref{s2:q-dim}, the linearised
vorticity equation in MHD is given by
\begin{equation}\label{app:MHD1}
	\left(\frac{\dy}{\dy t}+\Ubar\frac{\dy}{\dy x}\right)q=
	-\frac{\dy\Qbar}{\dy y}\frac{\dy\psi}{\dy x}+\Bbar\frac{\dy j}{\dy x}
	+\left(\Bbar\frac{\dy\Jbar}{\dy y}\right)\frac{\dy \eta}{\dy x}.
\end{equation}
For the analogous problem in the Bousinessq system, using the fact that
$(\dy/\dy t+\Ubar\dy/\dy x)b=-wN^{2}(z)$ and $(\dy/\dy t+\Ubar\dy/\dy
x)\zeta=w$, integrating yields the relation that $b=-\eta N^{2}(z)$, where
$N^{2}(z)$ is the background buoyancy frequency. The resulting vorticity
equation reads
\begin{equation}\label{app:q-dim}
	\left(\frac{\dy}{\dy t}+\Ubar\frac{\dy}{\dy x}\right)q=
	-\frac{\dy\Qbar}{\dy z}\frac{\dy\phi}{\dy x}-N^{2}\frac{\dy\zeta}{\dy x}.
\end{equation}
Then the formal similarity is that
$-\Bbar(\dy\Jbar/\dy y)\leftrightarrow N^{2}$. The analogy is not entirely
complete here because, as we have seen before, for $\Bbar(\dy\Jbar/\dy y)>0$,
the destabilising effect of $\Bbar(\dy\Jbar/\dy y)$ is counteracted somewhat by
the term $\Bbar(\dy j/\dy x)$. In the case where we assume interfacial waves or
plane waves, the analogy is augmented by relating $j$ to $A$ and then to $\eta$,
which then gives $\Bbar(\Bbar-\Delta\Jbar/2k)\leftrightarrow N^{2}$. However, as
discussed in the text, isolated $\delta$-function solution neglects
contributions away from the `jump', and predicts instability when the global
state (in the absence of a flow) should be stable \citep[e.g.,][]{Lundquist51}.
We note, for completeness, gravity wave propagation and Rayleigh--Taylor
instability has an analogous description in terms of waves and vorticity action,
as in figure~\ref{fig:CRW_eigenstructure2}, by replacing current anomalies with
buoyancy anomalies and considering the action of the resulting baroclinic
torque (see also figure~2 and 3 of \citealt{Harnik-et-al08}).

We note that, in the stratified problem, without assuming interfacial waves, the
evolution equation for the perturbation energy is given by
\begin{equation}
	\frac{1}{2}\frac{\dy}{\dy t}\left\langle(u^{2}+w^{2})
	+N^{2}\zeta^{2}\right\rangle=
	-\left\langle uw\frac{\dy\Ubar}{\dy z}\right\rangle
\end{equation}
where the angle bracket denotes a domain integral. So if $N^{2}<0$, it is still
possible for the perturbation kinetic energy to grow even in the absence of a
background flow. In the MHD case, a similar manipulation leads first to
\begin{equation}
	\frac{\dy u}{\dy t}+\Ubar\frac{\dy u}{\dy x}=
	-v\frac{\dy\Ubar}{\dy y}-\frac{\dy}{\dy x}(\cdots),\qquad
	\frac{\dy v}{\dy t}+\Ubar\frac{\dy v}{\dy x}=
	\Bbar\left(\frac{\dy\Jbar}{\dy y}\eta+j\right)-\frac{\dy}{\dy y}(\cdots),
\end{equation}
where the terms in $(\cdots)$ between the two equations are the same. Again, it
is the appearance of the perturbation current that makes the analogy incomplete.
If we however assume that $\dy\Jbar/\dy y=\Delta\Jbar\delta(y-L)$ and
$j=\jhat\ex^{\zi kx}\delta(y-L)$, then it may be shown that, at $y=L$,
\begin{equation}
	\frac{1}{2}\frac{\dy}{\dy t}\left\langle(u^{2}+v^{2})
	+\Bbar(2k\Bbar-\Delta\Jbar)\eta^{2}\right\rangle=
	-\left\langle uv\frac{\dy\Ubar}{\dy y}\right\rangle,
\end{equation}
and, in the absence of a background flow, if $\Bbar\Delta\Jbar>0$, it is still
possible for the perturbation kinetic energy to grow at the expense of magnetic
energy, resulting in instability.

With regards to the non-dimensional units, we recall that, in the stratified
case, the governing non-dimensional parameter is $Ri=N^{2}/(\dy\Ubar/\dy
z)^{2}$. In MHD, the governing non-dimensional number is
\begin{equation}
	M^{2}=\frac{B_{0}^{2}}{U_{0}^{2}}\sim
	\frac{\Bbar(L^{2}\Bbar)}{L^{2}\Ubar^{2}}\sim
	\frac{-\Bbar(\dy\Jbar/\dy y)}{(\dy\Ubar/\dy y)^{2}},
\end{equation}
and so $M^2$ has a formal analogy with $Ri$. We should stress that this is only
a formal analogy, since it does not take into account of the $\Bbar(\dy j/\dy
x)$ term, which results in wave modes that are fundamentally different from the
gravity wave modes. We note that, in a similar vein, \cite{Stern63} defines what
he calls the magnetic Richardson number as (translated into our notation)
$Ri_{m}=(\dy\Bbar/\dy y)^{2}/ (\dy\Ubar/\dy y)^{2}$.


\section{Formulation in terms of $q$ and $j$ variables}\label{appendix2}

One alternative approach is to use $q$ and $j$ as the fundamental variable over
$q$ and $\eta$ when considering interface solutions. One immediate issue is
that, from the current perturbation equation \eqref{s2:j-dim}, $\Qbar$ and
$\Jbar$ are not-well defined at the interface locations. A further approximation
that one might make is that the terms with the coefficient $\Qbar$ and $\Jbar$
are small compared to $\dy\Qbar/\dy y$ and $\dy\Jbar/\dy y$; this may be
appropriate if we have reason to believe that interfacial waves are the most
important aspect to the dynamics, perhaps when the basic state gradients are
strong. With this assumption, we obtain the system of equations
\begin{subequations}\label{app:MHD}\begin{align}
	\left(\frac{\dy}{\dy t}+\Ubar\frac{\dy}{\dy x}\right)q=&
	-\frac{\dy\Qbar}{\dy y}\frac{\dy\psi}{\dy x}+\Bbar\frac{\dy j}{\dy x}
	+\frac{\dy\Jbar}{\dy y}\frac{\dy A}{\dy x}, \label{app:MHD-q}\\
	\left(\frac{\dy}{\dy t}+\Ubar\frac{\dy}{\dy x}\right)j=&
	+\frac{\dy\Qbar}{\dy y}\frac{\dy A}{\dy x}+\Bbar\frac{\dy q}{\dy x}
	-\frac{\dy\Jbar}{\dy y}\frac{\dy\psi}{\dy x}.
\end{align}\end{subequations}
With this, we notice that, if we define a generalised streamfunction and
vorticity as $\phi^{\pm}=(\psi\pm A)/2$ and $\mathcal{Q}^{\pm}=(q\pm j)/2$, then
we have $\grad^{2}\phi^{\pm}=Q^{\pm}$ and that
\begin{equation}
	\left[\frac{\dy}{\dy t}+(\Ubar\mp\Bbar)\frac{\dy}{\dy x}\right]
	\mathcal{Q^{\pm}}=
	-\left(\frac{\dy\Qbar}{\dy y}\pm\frac{\dy\Jbar}{\dy y}\right)
	\frac{\dy\phi^{\mp}}{\dy x}.
\end{equation}
This has some formal similarities to employing Elsasser variables
\citep[e.g.,][]{Biskamp-MHD} in writing the MHD equations, although this is of
course different since we have made an approximation by dropping certain terms
in the $j$ equation.


\bibliographystyle{jfm}

\label{lastpage}

\end{document}